%% file: vdbosch.tex
\documentstyle{mn}
\input{psfig}
\input{macros_vdbosch}

\begin{document}


\title[The Impact of Cooling and Feedback on Disc Galaxies]
      {The Impact of Cooling and Feedback on Disc Galaxies}
\author[F.C. van den Bosch]
       {Frank C. van den Bosch \\
        Max-Planck Institut f\"ur Astrophysik, Karl Schwarzschild
         Str. 1, Postfach 1317, 85741 Garching, Germany}


\date{}

\pagerange{\pageref{firstpage}--\pageref{lastpage}}
\pubyear{2001}

\maketitle

\label{firstpage}


\begin{abstract}
  In  the  standard  picture  of  galaxy formation  mass  and  angular
  momentum are  the two main parameters that  determine the properties
  of  disc galaxies. The  details of  how the  gas inside  dark matter
  haloes is transformed into a luminous disc, however, depend strongly
  on the  physics of star  formation, feedback and cooling,  which are
  poorly understood. The efficiencies of these astrophysical processes
  are  ultimately responsible  for setting  the galaxy  mass fractions
  $f_{\rm gal}$ defined as the  ratio of galaxy mass (disc plus bulge)
  to  total virial  mass.  Therefore,  if we  could  somehow determine
  $f_{\rm  gal}(M_{\rm  vir})$   from  observations  of  the  luminous
  component of  disc galaxies,  this would allow  us to  put stringent
  constraints  on the  efficiencies  of cooling  and feedback.   This,
  however,  requires estimating  the  total virial  mass,  which is  a
  delicate problem.  In this  paper we use detailed, analytical models
  for the  formation of disc  galaxies to investigate the  impact that
  cooling  and  feedback  have  on their  structural  properties.   In
  particular, we investigate which observables extracted directly from
  the models  are best suited as  virial mass estimators,  and to what
  extent  they  allow  the  recovery  of the  model  input  parameters
  regarding the feedback and  cooling efficiencies.  Contrary to naive
  expectations,  the  luminosities  and  circular velocities  of  disc
  galaxies  are  extremely  poor  indicators  of  total  virial  mass.
  Instead, we show that the  product of disc scale length and rotation
  velocity  squared yields a  much more  robust virial  mass estimate,
  which allows a fairly accurate recovery of the galaxy mass fractions
  as function of  virial mass. We also show that  feedback can cause a
  narrow  correlation   between  $f_{\rm  gal}$  and   the  halo  spin
  parameter, as  recently found by  van den Bosch, Burkert  \& Swaters
  from  an  analysis of  dwarf  galaxy  rotation  curves.  Finally  we
  investigate the impact that cooling and feedback have on the colors,
  metallicities, star formation histories and Tully-Fisher relation of
  disc  galaxies. 
\end{abstract}




\section{Introduction}
\label{sec:intro}

Currently, the  main uncertainties in our picture  of galaxy formation
are related to the intricate processes of cooling, star formation, and
feedback. In particular, we  need to understand how efficient feedback
is in expelling  baryons from dark matter haloes,  and how it enriches
the  IGM.  This  in turn  influences the  cooling  efficiencies, which
influences the  subsequent star formation rates,  which influences the
subsequent amount of  energy that is fed back into  the IGM, etc.  The
importance of  this complicated loop of `gastrophysics'  has long been
realized.  Already  the first detailed  models of galaxy  formation by
White \&  Rees (1978) revealed  an important problem intrinsic  to any
hierarchical formation  scenario in which  small mass clumps  merge to
form larger and larger  structures.  At early times, collapsed objects
have much  higher densities, and therefore much  shorter cooling times
than at the  present time.  Consequently, when these  objects merge to
form larger  and larger structures (i.e., galaxies  and clusters), the
vast majority  of their  baryons have already  cooled, so that  by the
present  time  there  is  basically   no  gas  left  to  make  up  the
intergalactic   or  intra-cluster   medium  observed   throughout  the
Universe.  This problem has  become known as the `cooling catastrophe'
and  is  generally interpreted  as  a  requirement  for some  sort  of
feedback mechanism that can pump energy back into the gas to lower its
cooling  efficiency (see  Balogh \etal  2001  for an  overview of  the
current status).

In the past, two related  techniques have been used to investigate the
impact  of  cooling  and   feedback  on  galaxy  formation:  numerical
simulations  (e.g., Katz,  Weinberg  \& Hernquist  1996; Fardal  \etal
2001; Pearce \etal 2000; Kay \etal 2001), and semi-analytical modeling
(e.g.,  Kauffmann, White  \&  Guiderdoni 1993;  Somerville \&  Primack
1999; Cole \etal 2000).   In both cases phenomenological prescriptions
are used  to describe star  formation and feedback, and  the resulting
model  galaxies  are  compared  to observational  data.   However,  an
important problem is  that so far none of these  models have been able
to  successfully  fit  all  observables.   This  is  likely  to  be  a
consequence  of the  simplicity of  the  phenomenological descriptions
used.  An additional problem is that assumptions have to be made about
poorly constrained model ingredients  such as stellar populations, the
stellar  initial  mass  function,  dust extinction,  etc.   Often  the
ignorance  regarding  these  ingredients   is  hidden  in  free  model
parameters, which  have hampered the model's ability  to place direct,
stringent  constraints on  the efficiencies  of cooling  and feedback.
Finally,  the galaxy  models  are  poorly (in  the  case of  numerical
simulations)  or not  at  all  (in the  case  of most  semi-analytical
models) spatially  resolved, which can complicate  a direct comparison
with observations.

In this paper we therefore take a slightly different approach based on
new models for the formation  of disc galaxies that are both spatially
and temporally  resolved. Rather than  comparing the models  to actual
observations  of real  disc  galaxies (which  is  postponed to  future
papers), we  investigate how well observables  {\it extracted directly
from the models} can be used  to {\it recover} the input parameters of
the model  regarding the feedback and cooling  efficiencies.  This has
the advantage  that even though  the assumptions underlying  the model
are not necessarily correct,  and the phenomenological descriptions of
star formation and feedback  are certainly oversimplified, it provides
important insights regarding the ability to use actual observations to
constrain  the  poorly understood  astrophysical  processes of  galaxy
formation.  The main  goal of this paper is therefore  not to tune our
models  to best  fit data,  but to  investigate how  both  cooling and
feedback impact on the observable properties of the model galaxies.

The cooling  and feedback efficiencies are  ultimately responsible for
setting the galaxy mass fractions  $f_{\rm gal} = M_{\rm gal} / M_{\rm
vir}$.  Here  $M_{\rm gal}$  is the total  {\it baryonic} mass  of the
galaxy (stars plus  gas, excluding the hot gas  in the halo).  Ideally
one  would therefore like  to observationally  obtain some  measure of
$f_{\rm  gal}$ as function  of virial  mass.  The  main focus  in this
paper, therefore, is  to investigate how well one  can hope to recover
$f_{\rm gal}(M_{\rm vir})$ from  direct observations of the population
of galaxies.  Unfortunately, obtaining  a measure of $f_{\rm gal}$ for
a single galaxy faces two  important challenges.  First of all we need
to be able  to infer the total baryonic  mass from observations, which
requires a conversion of luminosities and HI and CO masses to stellar,
atomic   and   molecular   masses,  respectively.    The   appropriate
conversation  ratios, however,  are still  relatively  uncertain. Even
more  problematic  is  inferring   the  total  virial  mass  from  the
photometry  and kinematics  of the  luminous component.   An important
focus of this paper, therefore,  is to identify those observables that
are best suited as indicators of both $M_{\rm gal}$ and $M_{\rm vir}$.

After  introducing the  models (Section~\ref{sec:models}),  we discuss
one   particular   model   galaxy   in  detail   to   illustrate   the
characteristics   of  the  models   (Section~\ref{sec:fiducial}).   In
Section~\ref{sec:barfrac} we  explore how cooling  and feedback impact
on the galaxy  mass fractions, and we search  for the observables that
are  best   suited  to  recover   the  model  input   parameters.   In
Sections~\ref{sec:tf}  and~\ref{sec:colors} we  discuss the  impact of
cooling and  feedback on  the Tully-Fisher relation,  and on  the star
formation   rates,  colors   and  metallicities   of   disc  galaxies,
respectively. We summarize our results in Section~\ref{sec:concl}.

\section{The Models}
\label{sec:models}

The model for the formation of disc galaxies used here is presented in
van  den  Bosch (2001;  hereafter  paper~1).   Here  we give  a  short
overview  of the  main  ingredients of  the  model, and  we refer  the
interested reader to Paper~1 for a more detailed description.

The main assumptions that characterize the framework of our models are
the following: (i) dark matter  halos around disc galaxies grow by the
smooth accretion of  mass, (ii) in the absence  of cooling the baryons
have the  same distribution of mass  and angular momentum  as the dark
matter, and (iii) the baryons conserve their specific angular momentum
when they cool.

The two main ingredients that determine the formation and evolution of
a disc galaxy, therefore, are  (the evolution of) the mass and angular
momentum  of the  virialized  object; $M_{\rm  vir}(r,z)$ and  $J_{\rm
vir}(r,z)$\footnote{Throughout  this  paper,  $r$  and  $z$  refer  to
spherical  radius and redshift,  respectively.}.  We  characterize the
angular  momentum  of  the  protogalaxies by  the  dimensionless  spin
parameter $\lambda = J_{\rm  vir} \vert E_{\rm vir} \vert^{1/2} G^{-1}
M_{\rm vir}^{-5/2}$. Here $E_{\rm vir}$  is the halo's energy, and $G$
is the gravitational constant. We follow Firmani \& Avila-Reese (2000)
and  make the  additional  assumptions that  (iv)  the spin  parameter
$\lambda$ of a given galaxy is constant with time, (v) each mass shell
that virializes is in solid  body rotation, and (vi) the rotation axes
of all shells  are aligned.  Although neither of  these assumptions is
necessarily  accurate, it  was shown  in Paper~1  that they  result in
halos with  angular momentum profiles that are  in excellent agreement
with  the  high  resolution  $N$-body  simulations  of  Bullock  \etal
(2001b).

The main outline of the models is as follows.  We set up a radial grid
between  $r=0$ and the present day  virial radius  of the model galaxy
and we follow the formation and  evolution of the  disc galaxy using a
few hundred time steps. We  consider six mass components: dark matter,
hot  gas, disc mass  (both in stars and  in cold gas), bulge mass, and
mass ejected by outflows from the disc.  The dark matter, hot gas, and
bulge mass are assumed to be distributed  in spherical shells, whereas
the disc stars and cold gas are assumed to  be in infinitesimally thin
annuli.  Each time step we  compute the changes  in these various mass
components in each radial bin, using the prescriptions detailed below.

\subsection{The evolution of the dark matter component}
\label{sec:mah}

The backbone of the models is  the formation and evolution of the dark
matter haloes, which is determined by the parameters of the background
cosmological model  and by  the power spectrum  $P(k)$ of  the initial
density  fluctuations. In  this  paper we  restrict  ourselves to  the
currently    popular   $\Lambda$CDM    model    with   $\Omega_0=0.3$,
$\Omega_{\Lambda}=0.7$, and  $H_0 = 100  h \kmsmpc$ with  $h=0.7$.  We
adopt the standard  CDM power spectrum normalized to  $\sigma_8 = 1.0$
and  we use  a baryon  density of  $\Omega_{\rm bar}  =  0.019 h^{-2}$
(Tytler \etal 1999).

\begin{figure}
\psfig{figure=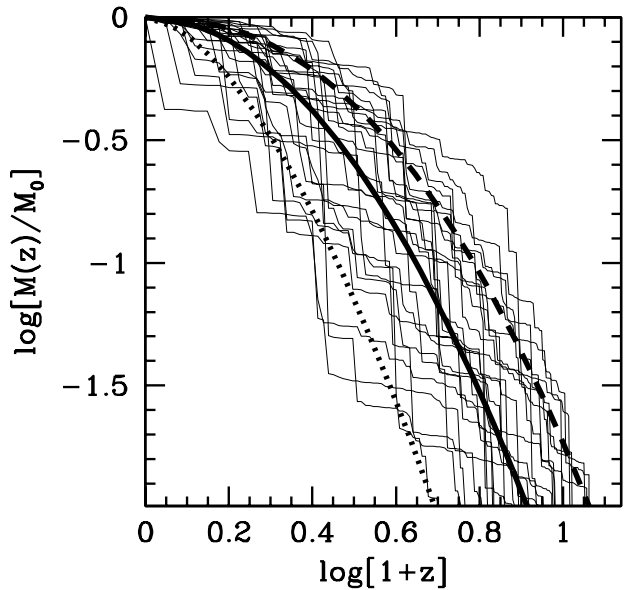,width=\hssize}
\caption{Thin lines correspond  to 25 randomly selected MAHs  for a $5
\times  10^{11}  h^{-1}  \Msun$  halo  in  our  standard  $\Lambda$CDM
Universe,  constructed using  the method  described in  van  den Bosch
(2002).   The  thick solid  line  corresponds  to  the universal  MAH
(equation~[\ref{unimah}]), and provides an accurate description of the
average  MAH for haloes  of this  mass.  The  dashed and  dotted lines
correspond  to  the  early  and   late  MAHs  described  in  the  text
(Section~\ref{sec:lambdaMAH}),  respectively, and roughly  outline the
distribution of MAHs for a  halo of this mass.  They are parameterized
by  the universal  MAHs for  halo masses  of $2  \times  10^{9} h^{-1}
\Msun$ and $10^{14} h^{-1} \Msun$ respectively.}
\label{fig:mah}
\end{figure}

The rate at which dark matter haloes grow in mass depends on cosmology
and  halo mass,  and can  be computed  using  extended Press-Schechter
theory (Bond  \etal 1991; Lacey \&  Cole 1993; Sheth  \& Tormen 1999).
In  Figure~\ref{fig:mah}  we  plot  25  random  realizations  of  mass
accretion histories  (hereafter MAH) of  a halo with present  day mass
$M_{\rm vir}(0) =  5 \times 10^{11} h^{-1} \Msun$  in the $\Lambda$CDM
cosmology adopted here (thin  lines). These MAHs are constructed using
the method  outlined in  van den Bosch  (2002). The thick  solid line
corresponds to  the average MAH of  a $5 \times  10^{11} h^{-1} \Msun$
halo (averaged over a large ensemble of MAHs), which is well fitted by
the universal form:
\begin{equation}
\label{unimah}
{\rm  log}\left({M_{\rm vir}(z) \over  M_{\rm vir}(0)}\right)  = -0.301
\left[ {{\rm log}(1+z) \over {\rm log}(1+z_f)} \right]^{\nu}
\end{equation}
(van den  Bosch 2002).  Here $M_{\rm  vir}(z)$ is the  virial mass at
redshift $z$, $z_f$ is the redshift at which the halo mass is half the
present day mass, and
\begin{eqnarray}
\label{nuB}
\lefteqn{\nu  =  1.211  +  1.858   \,  {\rm  log}[1+z_f]  +  0.308  \,
\Omega_{\Lambda}^{2} -  } \nonumber \\  & & 0.032 \,  {\rm log}[M_{\rm
vir}(0)/(10^{11} h^{-1} \Msun)]
\end{eqnarray}
The `formation' redshift $z_f$ depends  on halo mass and cosmology and
is easily  computed using the Press-Schechter  formalism (see Appendix
in van den Bosch 2002).  Throughout this paper we use this `universal'
MAH, and which is valid for all  halo masses and for a wide variety of
cosmologies, to describe the rate  at which dark matter haloes grow in
mass.  Each model galaxy is  therefore to be regarded as averaged over
all its possible MAHs.

We assume that  the dark matter virializes such  that at each redshift
the  halo is  spherical  with a  NFW  (Navarro, Frenk  \& White  1997)
density distribution:
\begin{equation}
\label{rhodm}
\rho_{\rm vir}(r) = \rho_s \left( {r \over r_s}\right)^{-1} 
\left( 1 + {r \over r_s}\right)^{-2}.
\end{equation}
We parameterize each dark matter halo by its total virial mass $M_{\rm
vir}$ and its concentration parameter $c = r_{\rm vir}/r_s$.  Here the
virial radius $r_{\rm  vir}$ is defined as the  radius inside of which
the  average halo  density is  $\Delta_{\rm vir}$  times  the critical
density  for  closure. For  the  $\Lambda$CDM  cosmology adopted  here
$\Delta_{\rm vir}  \simeq 101$  (Bryan \& Norman  1998). For  the halo
concentration parameter we adopt the model of Bullock \etal (2001c):
\begin{equation}
\label{ccc}
c(M_{\rm vir},z) = 4.0 \left( {1 + z_{\rm coll} \over 1 + z} \right)
\end{equation}
with $z_{\rm coll}$  the redshift at which a halo  with one percent of
the present day virial mass collapses.

Finally we consider a distribution of halo spin parameters given by
\begin{equation}
\label{spindistr}
p(\lambda){\rm d} \lambda = {1 \over \sigma_{\lambda} \sqrt{2 \pi}}
\exp\biggl(- {{\rm ln}^2(\lambda/\bar{\lambda}) \over 2
  \sigma^2_{\lambda}}\biggr) {{\rm d} \lambda \over
  \lambda},
\end{equation}
with  $\bar{\lambda}  =  0.06$  and  $\sigma_{\lambda}  =  0.6$.  This
log-normal distribution accurately  fits the distribution of $\lambda$
for haloes  in $N$-body simulations (e.g., Barnes  \& Efstathiou 1987;
Ryden 1988; Cole \& Lacey 1996; Warren \etal 1992).

\subsection{The formation of disc and bulge}
\label{sec:disc}

In  order to  compute  the formation  and  evolution of  the discs  we
proceed  as follows.   Each  time step  $\Delta  t$ a  new mass  shell
virializes. The mass  and angular momentum of that  shell are computed
from  the MAH  and the  requirement  that the  spin parameter  remains
constant,  respectively.   A   fraction  $f_{\rm  bar}  =  \Omega_{\rm
bar}/\Omega_0$ of this mass is in baryons, and is heated to the halo's
virial temperature.  The baryons dissipate energy radiatively, but are
assumed to  conserve their specific angular momentum.   The time scale
on which  they reach centrifugal equilibrium  in the disc  is given by
$t_c \equiv  {\rm max}[t_{\rm ff},  t_{\rm cool}]$, with  $t_{\rm ff}$
and $t_{\rm cool}$ the  free-fall and cooling times, respectively (see
paper~1  for   the  details).   We  use   the  collisional  ionization
equilibrium cooling functions of Sutherland \& Dopita (1993), assuming
a Helium  mass abundance  of $0.25$. The  metallicity of the  hot gas,
$Z_{\rm hot}$, is considered a free model parameter.

Self-gravitating   discs   tend   to   be  unstable   against   global
instabilities such as  bar formation.  Here we follow  the approach of
van  den Bosch  (1998, 2000)  and  Avila-Reese \&  Firmani (2000)  and
assume that an unstable disc transforms part of its disc material into
a bulge  component in  a self-regulating fashion  such that  the final
disc is marginally stable.  We consider the disc to be unstable if
\begin{equation}
\label{stabalpha}
\alpha_{\rm max} = \max_{0 \leq r \leq r_{\rm vir}} \left(
{V_{\rm disc}(r) \over V_{\rm circ}(r)}\right) < 0.7.
\end{equation}
(Christodoulou, Shlosman \& Tohline  1995). Here $V_{\rm disc}(r)$ and
$V_{\rm circ}(r)$  are the circular  velocities of the disc  (cold gas
plus  stars) and the  composite disc-bulge-halo  system, respectively.
Throughout we assume  that the disc is infinitesimally  thin, and each
time step we use the adiabatic invariant formalism of Blumenthal \etal
(1986) and   Flores  \etal   (1993)  to   compute   the  gravitational
contraction of the dark matter  induced by the baryons settling in the
disc.
\begin{figure*}
\centerline{\psfig{figure=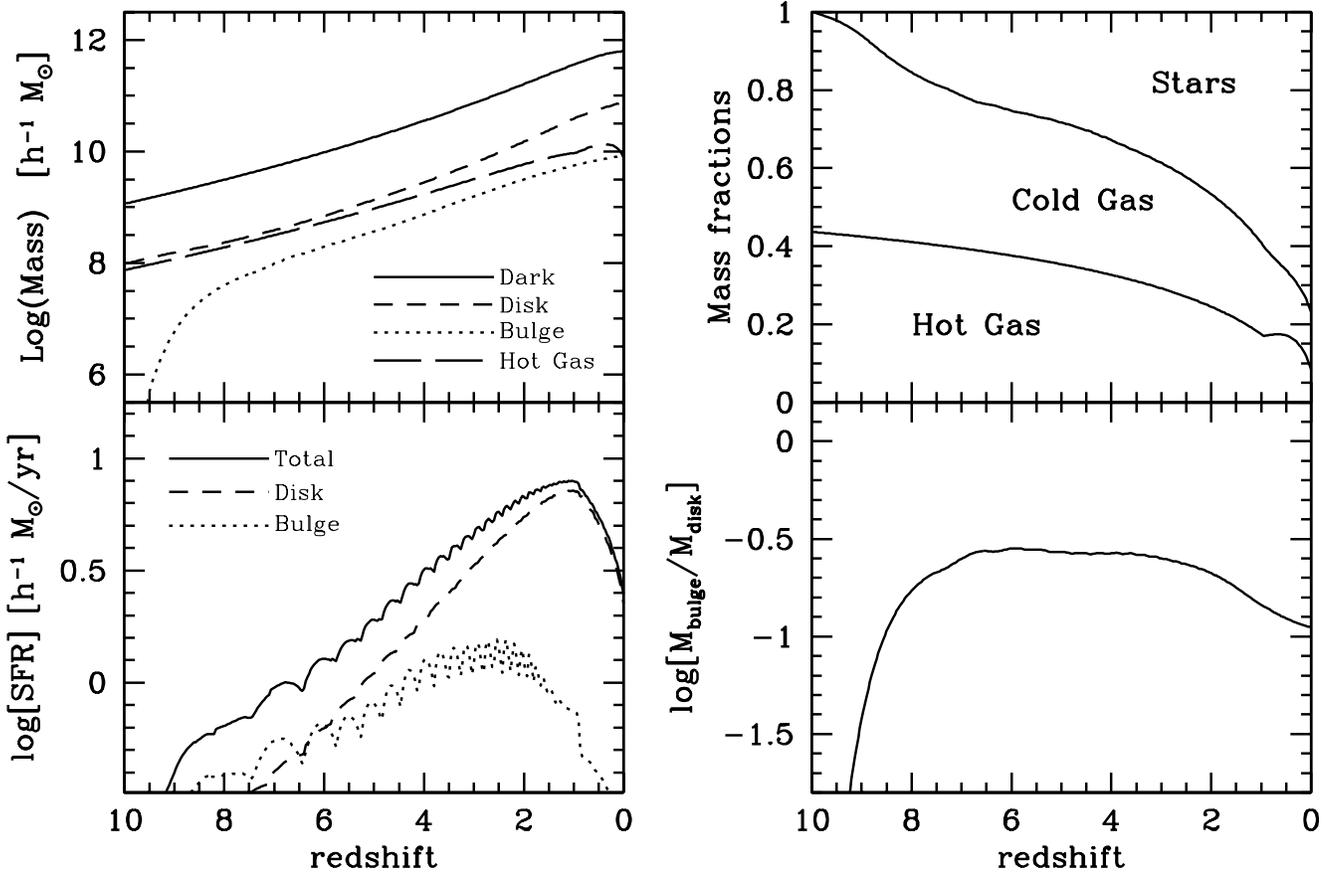,width=\hdsize}}
\caption{Some  results  for our  fiducial  model  galaxy discussed  in
Section~\ref{sec:fiducial}.   The  upper  left  panel plots  the  mass
evolution of various  components, as indicated. In the  upper panel on
the right,  the cumulative baryonic mass fractions  are plotted, which
show how initially all the baryons are in the gas phase (either hot or
cold),  which is  rapidly converted  into stars.   The  star formation
histories of the  disc and bulge are plotted in  the lower left panel,
while the  lower right  panel shows how  the bulge-to-disc  mass ratio
evolves with time.}
\label{fig:fiducial}
\end{figure*}

\subsection{Star formation and Feedback}
\label{sec:sf}

For the star formation rate we adopt the standard Schmidt (1959) law:
\begin{equation}
\label{Schmidt_law}
\psi(R) = 2.5 \times 10^{-4}\Msun  {\rm yr}^{-1} {\rm kpc}^{-2} \, \left(
{\Sigma_{\rm gas}(R) \over 1.0 \Msun {\rm pc}^{-2}} \right)^{1.4}
\end{equation}
where the  numerical values are  determined observationally (Kennicutt
1998).  This simple empirical law  holds over many orders of magnitude
in gas  surface density, and even applies  to circum-nuclear starburst
regions.   However, when  applied to  {\it local}  gas  densities, the
Schmidt law breaks down at  large disc radii, where the star formation
is found  to be  abruptly suppressed.  In  a seminal  paper, Kennicutt
(1989) showed that these  radii correspond to the radii  where the gas
surface  density falls  below the  critical surface  density  given by
Toomre's (1964) stability criterion:
\begin{equation}
\label{Toomre}
\Sigma_{\rm crit}(R) = {\sigma_{\rm gas} \, \kappa(R) 
\over 3.36  \, G \, Q}.
\end{equation}
Here $Q$ is a dimensionless constant near unity, $\sigma_{\rm gas}$ is
the  velocity dispersion  of the  gas,  and $\kappa$  is the  epicycle
frequency.  Each  time step, and for  each radial bin,  we compute the
mass  in cold  gas that  is transformed  into stars  by  solving ${\rm
d}\Sigma_{\rm  gas}/{\rm  d}t  =   -\psi$  with  the  constraint  that
$\Sigma_{\rm   gas}$   can   not   be  depleted   below   $\Sigma_{\rm
crit}$. Throughout  we adopt $Q=1.5$  and $\sigma_{\rm gas} =  6 \kms$
(Kennicutt 1989). The gas that is transferred into the bulge component
is assumed to form stars instantaneously with 100 percent efficiency.

In addition to the simple Schmidt law, we have also experimented with
a star formation rate given by
\begin{equation}
\label{dynamicalSFR}
\psi(R) = 0.017 \Sigma_{\rm gas}(R) {V_c(R) \over R}
\end{equation}
(Silk  1997),  again  combined   with  the  star  formation  threshold
criterion.  With this parameterization each orbit $\sim 10$ percent of
the  available  cold gas  is  transformed  into  stars.  As  shown  by
Kennicutt (1989), this  description yields an equally good  fit to the
data as the Schmidt-law  of equation~(\ref{Schmidt_law}).  As it turns
out,  our models  yield virtually  indistinguishable results  for both
SFRs, and we  therefore adhere to the Schmidt-law  description in what
follows.

When stars evolve  they put energy  into the interstellar medium (ISM)
which impacts on the further evolution of  the galaxy. By resorting to
an  empirical description  of  the star formation,  we are  implicitly
taking account of the  effects that these  feedback processes have  on
the star formation rate.  What is not taken  into account, however, is
a possible feedback-driven outflow of gas from the  disc.  Here we use
a simple  parametric  model,  similar to  the   ones  used in  various
semi-analytical models for galaxy formation. We assume that the amount
of gas blown out of the disc is proportional to the total energy input
by supernovae (SNe) and  inversely proportional to the escape velocity
squared. At each radial bin, the cold gas mass that is ejected is
given by
\begin{equation}
\label{mass_eject}
\Delta M_{\rm eject} =  {\varepsilon_{\rm fb} \, 
\eta_{\rm  SN} \, E_{\rm SN} \over V_{\rm esc}^2} \, \Delta M_{*}
\end{equation}
(cf.   Kauffmann \etal  1993;  Natarajan 1999).   Here  $E_{\rm SN}  =
10^{51}$ergs is the energy produced  by one SN, $\eta_{\rm SN}$ is the
number of  SNe per solar  mass of stars  formed, $V_{\rm esc}$  is the
local escape velocity, and  $\varepsilon_{\rm fb}$ is a free parameter
that  describes  what  fraction  of  the energy  released  by  SNe  is
converted into kinetic energy to drive the outflow. For simplicity, we
assume  that the ejected  mass is  forever lost  from the  system: the
ejected mass is not considered for later infall, and the corresponding
metals are not used to enrich the infalling gas.

\subsection{Stellar population modeling \& chemical evolution}
\label{sec:stelpop}

In order  to convert the stellar  masses into luminosities  we use the
latest  version of the  Bruzual \&  Charlot (1993)  stellar population
synthesis models.  These models provide  the luminosities of  a single
burst stellar population as function of age $t$ and metallicity $Z$ in
various optical passbands.  In  order to model the chemical enrichment
of   the  ISM   we   follow  the   standard  instantaneous   recycling
approximation (IRA). We assume that  a fraction ${\cal R}$ of the mass
in stars formed is instantaneously returned to the cold gas phase with
a yield $y$  (which is defined as the fraction  of mass converted into
stars  that is  returned to  the  ISM in  the form  of newly  produced
metals).

 In  each disc  annulus,  and  at each  time  step, mass  conservation
implies
\begin{equation}
\label{mcold}
\Delta M_{\rm cold} = \Delta M_{\rm cool} - (1-{\cal R}) \Delta M_{*} -
\Delta M_{\rm eject}
\end{equation}
and for the mass in metals one thus obtains
\begin{eqnarray}
\label{mmetal}
\Delta M_{\rm metal} & = & Z_{\rm hot} \Delta M_{\rm cool} - 
Z_{\rm cold}  \Delta M_{\rm eject} - \nonumber \\
 & & Z_{\rm  cold} (1-{\cal R}) \Delta M_{*} + y \Delta M_{*}
\end{eqnarray}
Here $\Delta M_{\rm  cool}$ is the mass that, in  the given time step,
has cooled,  and $\Delta M_{*}$ is  the mass that  is transformed into
stars.   We use  these two  equations to  track the  evolution  of the
metallicity of the  cold gas in the disc,  $Z_{\rm cold}$, as function
of both  time and radius.  Throughout  we adopt the  Scalo (1986) IMF,
for which $\eta_{\rm SN} =  4 \times 10^{-3} \Msun^{-1}$ and ${\cal R}
= 0.25$.  The stellar yield $y$, finally, is left a free parameter.

\subsection{Missing Ingredients}
\label{sec:missing}

\begin{figure*}
\centerline{\psfig{figure=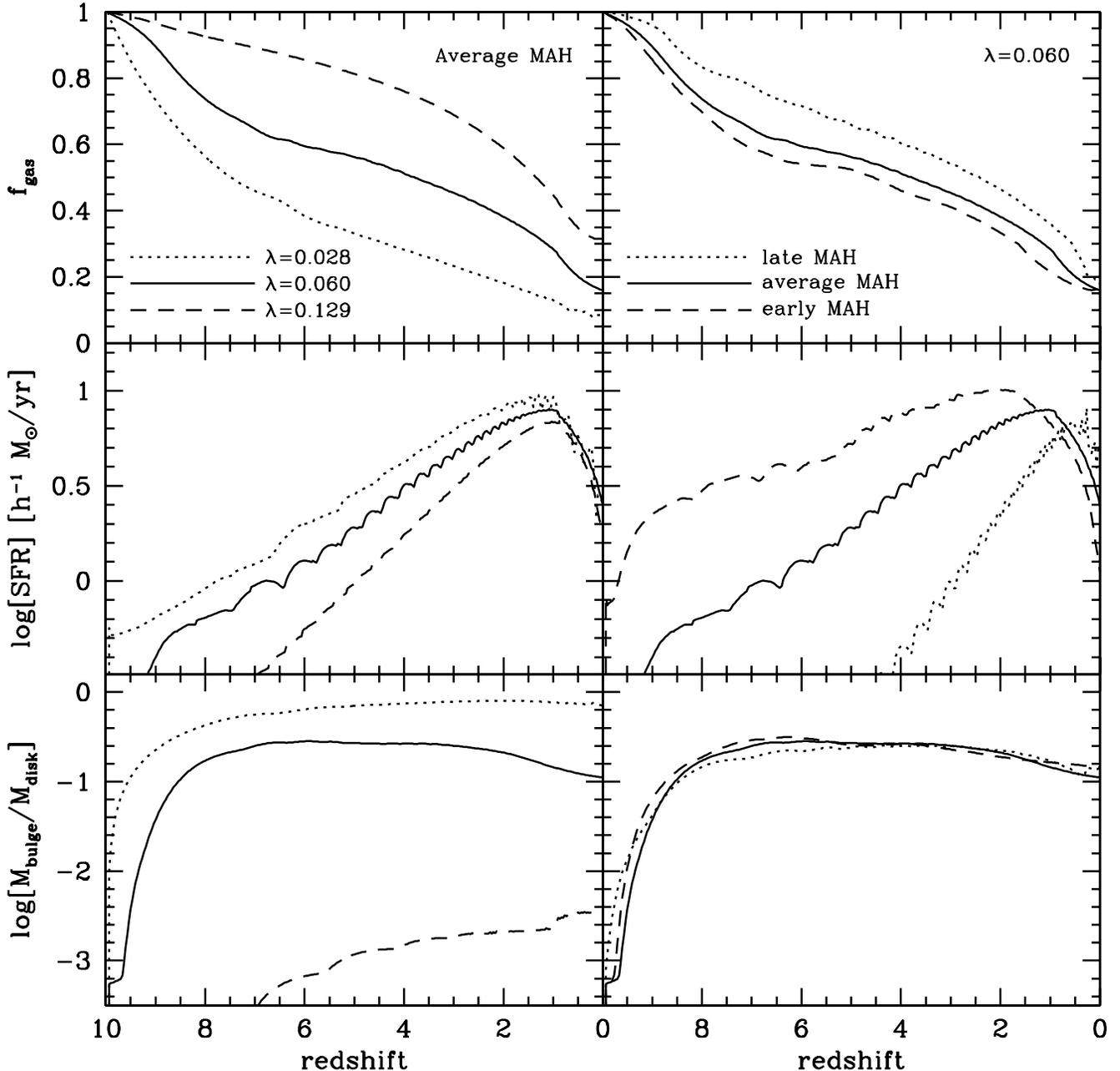,width=\hdsize}}
\caption{The influence of changes  in spin parameter (left panels) and
MAH (right panels) on the  outcome of our fiducial model galaxy.  From
the top to bottom we plot the disc gas mass fractions, the total (disc
plus bulge)  star formation rates, and the  bulge-to-disc mass ratios,
all  as  functions of  redshift.  Haloes  with  less angular  momentum
produce  more  bulge-dominated  disc  galaxies  with  lower  gas  mass
fractions.  The MAH  mainly sets  the  star formation  history of  the
galaxies,  but   does  not  significantly   influence  the  structural
parameters.}
\label{fig:varfiduc}
\end{figure*}

It is important  to understand the various shortcomings  of the models
discussed  above. First  of all,  for each  galaxy we  adopt  the {\it
average} mass accretion  history for systems of that  mass. In reality
there    is    a   relatively    large    scatter    in   MAHs    (see
Figure~\ref{fig:mah}), which introduces some additional scatter in for
instance the  star formation rates, and  thus the colors  of our model
galaxies (cf. right panels of Figure~\ref{fig:varfiduc}). Furthermore,
in our  models all halos of  a given mass have  the same concentration
parameter, whereas  in reality dark  matter halos reveal  a relatively
large amount of scatter in  $c(M_{\rm vir})$ (Jing 2000; Bullock \etal
2001c).   In fact,  Wechsler \etal  (2001)  have shown  that the  halo
concentration parameter  is strongly correlated with the  MAH. None of
these effects are taken into  account here. Also, by assuming that all
mass is accreted in a smooth  fashion we ignore the effect of discrete
mergers.  Although  the fragility of  disks (T\'oth \&  Ostriker 1992)
suggests that mergers can not have played a dominant role during their
formation, some  amount of merging almost certainly  occurs.  Since we
ignore  all these  effects, each  model galaxy  should be  regarded as
averaged over  all its possible MAHs, which  underestimates the amount
of scatter in  various properties of our model  galaxies.  However, we
are not  attempting to  compare our models  to data. Instead  our main
goal is  to investigate the impact  that cooling and  feedback have on
the  models,  and  underestimating  the  scatter  will  only  help  to
highlight these effects.

An important oversimplification of our  models is the treatment of the
galactic  winds. We  assume that  any material  expelled  by supernova
feedback  escapes the  halo without  enriching the  hot halo  gas.  In
reality some of the outflowing  material may remain bound to the halo,
enrich the hot gas (which increases its cooling efficiency), and later
cool back onto  the disk. However, very little is  known about how the
outflowing,  enriched  material  interacts  with  the  hot  halo  gas.
Therefore, rather than introducing  some additional free parameters to
our  models,  we  have  simply  chosen a  (perhaps  somewhat  extreme)
prescription. As emphasized above, our main goal is to investigate how
we may hope to infer $f_{\rm gal}(M_{\rm vir})$ from observations. For
this purpose,  the models  do not necessarily  have to yield  the most
physical/accurate estimate of the true $f_{\rm gal}(M_{\rm vir})$.

\section{A fiducial model galaxy}
\label{sec:fiducial}

Before  we investigate  large samples  of model  disc galaxies,  it is
useful to investigate  one particular model galaxy in  more detail. We
exclude feedback for the moment (i.e., $\varepsilon_{\rm fb} = 0$) and
assume that  baryons that  enter the virial  radius have  already been
enriched to one-third Solar metallicity ($Z_{\rm hot} = Z_{\odot}/3$).
A particular model  galaxy is parameterized by its  present day virial
mass $M_{\rm vir}(0)$ and its spin parameter $\lambda$.  We focus on a
model galaxy with $M_{\rm vir}(0) = 5 \times 10^{11} h^{-1} \Msun$ and
$\lambda=0.06$.   The   MAH  is  given   by  the  Universal   form  of
equation~(\ref{unimah}).   These parameters result  in a  model galaxy
that  is fairly  similar to  the Milky  Way (MW):  at $z=0$  the model
galaxy has  a disc scale length  of $4.5$ kpc, a  rotation velocity at
$8.5$ kpc of $230 \kms$, and  a bulge-to-disc mass ratio of $0.1$.  We
tune the yield to $y=0.01$ so that  the cold gas at $8.5$ kpc from the
center has solar metallicity.

The upper left panel  of Figure~\ref{fig:fiducial} plots the evolution
of the various mass components  with time (redshift).  The upper panel
on the right plots the cumulative mass fractions of the baryons inside
the virial  radius that at each redshift  are in the form  of hot gas,
cold gas  (in the disc), and stars  (in both the disc  and bulge).  At
$z=10$, when we start our  model computations, about 50 percent of the
baryonic mass inside the virial radius  is in the hot gas phase, while
the rest has already had time to cool. Star formation immediately sets
in, transferring  more and more of  the cold gas into  stars, while at
the same time  new baryons enter the virial radius,  are heated to the
virial temperature,  and cool to  become part of  the cold gas  in the
disc. At  $z=0$ about 80  percent of all  the baryonic mass is  in the
form of stars, while the remaining 20 percent is roughly equally split
amongst  cold  and   hot  gas.   The  lower  right   panel  plots  the
bulge-to-disc mass ratio ($B/D$) as  function of time. At $z \simeq 9$
the  disc  becomes  unstable  and  a bulge  starts  to  form,  rapidly
increasing  $B/D$ to about  $0.3$.  At  later times  ($z \lta  2$) the
bulge-to-disc ratio slowly  decreases again to a present  day value of
$\sim     0.1$.     Finally,     the    lower     left     panel    of
Figure~\ref{fig:fiducial}  plots the  star formation  rates  (SFRs) of
both the disc  and bulge. At high redshifts ($z \gta  5$) the disc and
bulge have very  comparable SFRs. At lower redshifts  the disc clearly
dominates the  total SFR, which  peaks at $z  \sim 1$, after  which is
declines  rapidly  to  a  present  day  SFR  of  $\sim  3  \Msun  {\rm
yr}^{-1}$.  The  `wobbly' behavior  of  the SFR  of  the  bulge is  an
artifact  of the  way bulge  formation is  taken into  account  in the
models, but does not influence any of our results.

\subsection{Model dependence on spin parameter and mass accretion
history}
\label{sec:lambdaMAH}

In order  to gauge the  dependence of the  model galaxies on  the halo
spin  parameter we  compare  model galaxies  with  $\lambda =  0.028$,
$0.06$, and $0.129$.   These values correspond to the  $10$, $50$, and
$90$     percentile     points      of     the     distribution     of
equation~(\ref{spindistr})    with   $\bar{\lambda}   =    0.06$   and
$\sigma_{\lambda} =  0.6$. In addition  to mass and  angular momentum,
the  actual mass  accretion history  of a  dark matter  halo  can also
impact  on the final  outcome of  the model  galaxy. Although  for the
remainder   of   this   paper   we   adopt  the   universal   MAH   of
equation~(\ref{unimah}), which  corresponds to the {\it  average} of a
large ensemble of possible MAHs, it is important to understand how the
models depend  on variations of the  MAH with respect  to the average.
To that  extent we  construct three models  (with $M_{\rm vir}(0)  = 5
\times 10^{11}  h^{-1} \Msun$ and $\lambda=0.06$) that  only differ in
their  MAHs.   These  MAHs  are plotted  in  Figure~\ref{fig:mah}:  in
addition to  the average MAH (thick  solid line) we  consider an early
(dashed line) and a late  (dotted line) MAH, which roughly outline the
distribution  of possible MAHs  for a  halo with  $M_{\rm vir}(0)  = 5
\times 10^{11} h^{-1} \Msun$.
\begin{figure*}
\centerline{\psfig{figure=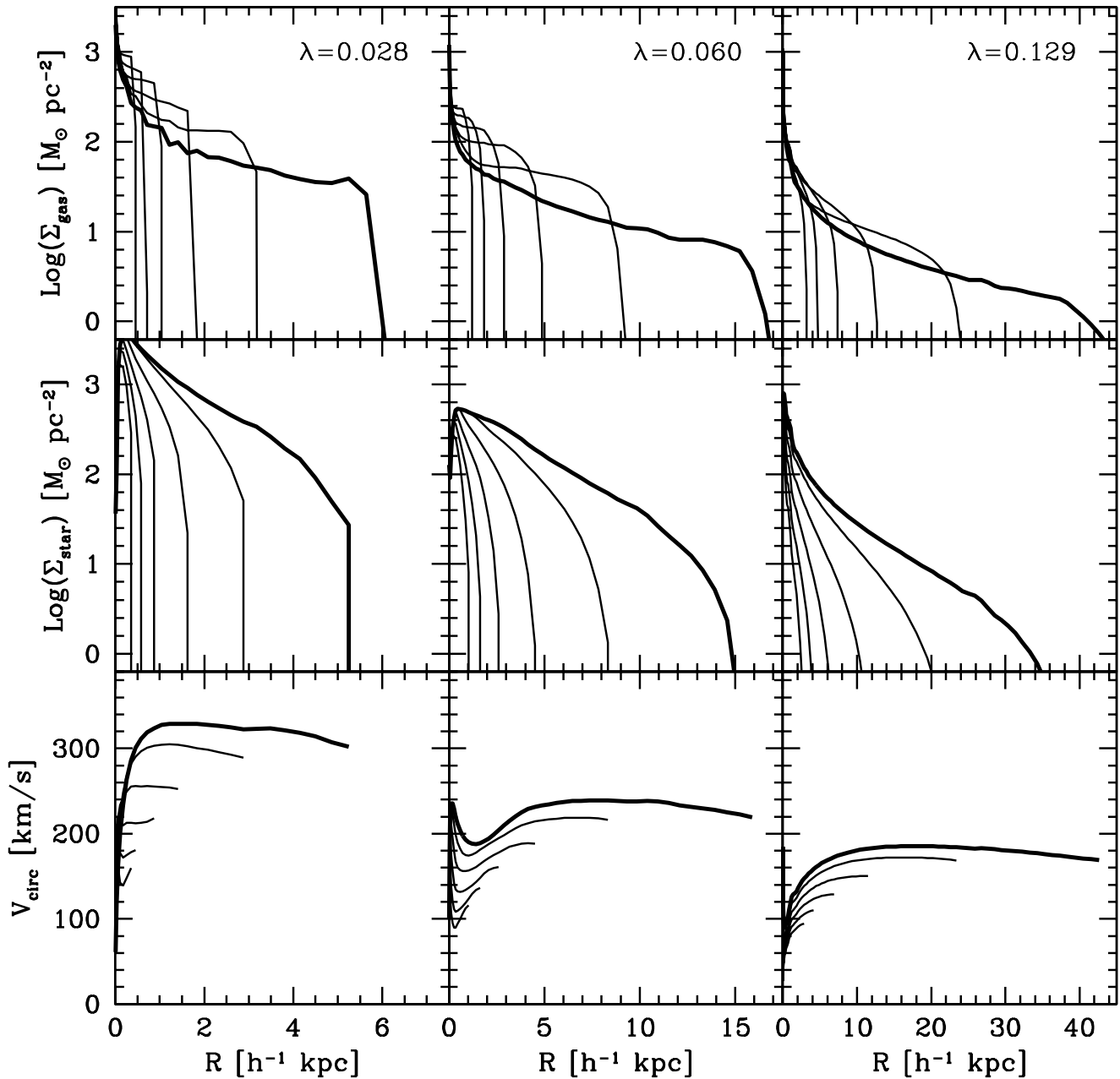,width=\hdsize}}
\caption{The  evolution of  the  surface densities  of  cold gas  (top
panels), stars  (middle panels), and rotation  curves (bottom panels),
all  as  function of  galactocentric  radius,  for  three models  with
$M_{\rm vir} = 5 \times  10^{11} h^{-1} \Msun$ and different halo spin
parameters  (as  indicated  in   the  top  panels).  The  thick  lines
correspond to  the present day  ($z=0$), while the various  thin lines
correspond to $z=5,4,3,2,1$. Note the inside-out character of the disc
formation, the  close to exponential surface  density distributions of
the stars, and the flat rotation curves.}
\label{fig:sdall}
\end{figure*}

In Figure~\ref{fig:varfiduc} we show the effects that a change in spin
parameter (left panels) and MAH  (right panels) have on the outcome of
the fiducial  model presented in  Figure~\ref{fig:fiducial}.  The halo
spin  parameter   determines  mainly  the   bulge-to-disc  ratio:  the
high-$\lambda$ model is  virtually bulgeless and with a  disc gas mass
fraction $f_{\rm gas}$  about twice as high as  for the $\lambda=0.06$
model.   Here $f_{\rm gas}  = M_{\rm  cold} /  (M_{\rm cold}  + M_{\rm
star})$ with $M_{\rm cold}$ and $M_{\rm star}$ the disc masses in cold
gas  and stars,  respectively.   The low  angular  momentum halo  with
$\lambda=0.028$ results in a galaxy for which the bulge mass is almost
equal  to  that  of the  disc  (i.e.,  the  resulting galaxy  is  more
reminiscent  of  an S0  than  a  spiral  galaxy).  Although  the  spin
parameter has a  mild influence on the SFR at $z  \gta 1$, the present
day SFR is virtually independent of $\lambda$: although systems with a
lower value of $\lambda$ have  a higher surface density, and therefore
a  higher star  formation  rate {\it  per  unit area},  they are  also
smaller,  such that  the  total, area-integrated  star formation  rate
depends only weakly on $\lambda$. The MAH, on the other hand, strongly
influences the SFR,  even at $z=0$, suggesting that  the rate at which
stars form is directly linked to the rate at which the galaxy accretes
mass.   The disc  gas mass  fraction  and bulge-to-disc  ratio do  not
depend strongly on the MAH, especially  not at $z=0$. Note that in our
models bulges  form only as  a consequence of disk  instabilities.  In
reality, part of the bulge may also form out of mergers of sub-clumps,
in which case the bulge-to-disk  ratio may depend more strongly on the
MAH than is the case here.

In Figure~\ref{fig:sdall}  we plot the  surface densities of  the cold
gas  (upper  row)  and  disc   stars  (middle  row)  as  functions  of
galactocentric radius at six different redshifts: $z=5,4,3,2,1,0$ with
the latter one plotted as  thick solid lines.  Results are plotted for
three different values of the  halo spin parameter as indicated in the
top panels.  All these models  have $M_{\rm vir}(0) = 5 \times 10^{11}
h^{-1} \Msun$ and  an average MAH.  The evolution  of the disc surface
densities with  redshift clearly illustrates the  inside-out growth of
the disc;  already at high  redshift the central surface  densities of
the stellar  discs are established, while  at $z=1$ the  disc is still
only about  half the present day  size.  At each  redshift the stellar
disc  nicely follows  an  exponential profile  with  a distinct  outer
truncation radius\footnote{An  exception is the  $z=0$ stellar surface
density profile of the  $\lambda=0.129$ model, which is more centrally
concentrated than  an exponential.  This  indicates a problem  for the
models which is discussed in  detail in Paper~1}.  The surface density
distribution of the cold gas is much shallower than that of the stars,
in good agreement  with observations. See paper~1 for  a more detailed
discussion of why the gas and cold gas follow distinct surface density
distributions.

The lower panels of  Figure~\ref{fig:sdall} plot the circular velocity
curves of  the model  galaxies (plotted out  to the maximum  radius at
which cold gas  is available, to mimic what an  observer might be able
to measure) at the same  redshifts.  The present day rotation curve of
the  $\lambda=0.06$  model  (middle  column)  peaks  at  small  radii,
reflecting the presence  of the bulge, reaches a  minimum of $\sim 200
\kms$ at about  $2$ kpc, and becomes fairly flat  at larger radii with
$V_{\rm circ}  \sim 230 \kms$.   All these features are  in remarkably
good agreement  with the  MW rotation curve  (Burton \&  Gordon 1978).
The rotation  curves for  the other two  models are  more featureless,
lacking  any  obvious transition  region  from  either  the bulge-  or
halo-dominated region to the disc-dominated region. The lack of such a
feature, often  referred to as  the `disc-halo' conspiracy  (e.g., van
Albada \etal  1985) is in good  agreement with data,  which shows that
there is nothing conspicuous about it.
\begin{table}
\begin{minipage}{\hssize}
\caption{Overview of Model Parameters.}
\label{tab:models}
\begin{tabular}{rclc}
Model & $Z_{\rm hot}/Z_{\odot}$ & $\varepsilon_{\rm fb}$ & $y$ \\
 PE-NFB &  $0.3$ & $0.0$  & $0.010$ \\
NPE-NFB &  $0.0$ & $0.0$  & $0.015$ \\
 PE-LFB &  $0.3$ & $0.02$ & $0.015$ \\
 PE-MFB &  $0.3$ & $0.05$ & $0.025$ \\
 PE-HFB &  $0.3$ & $0.10$ & $0.040$ \\
NPE-LFB &  $0.0$ & $0.02$ & $0.024$ \\
\end{tabular}

\medskip

Column~(1)  lists   the  model  ID.   Columns~(2)  --   (4)  give  the
metallicity  of the  hot  gas  (in units  of  solar metallicity),  the
feedback efficiency, and the stellar yield, respectively.

\end{minipage}
\end{table}
\begin{figure*}
\centerline{\psfig{figure=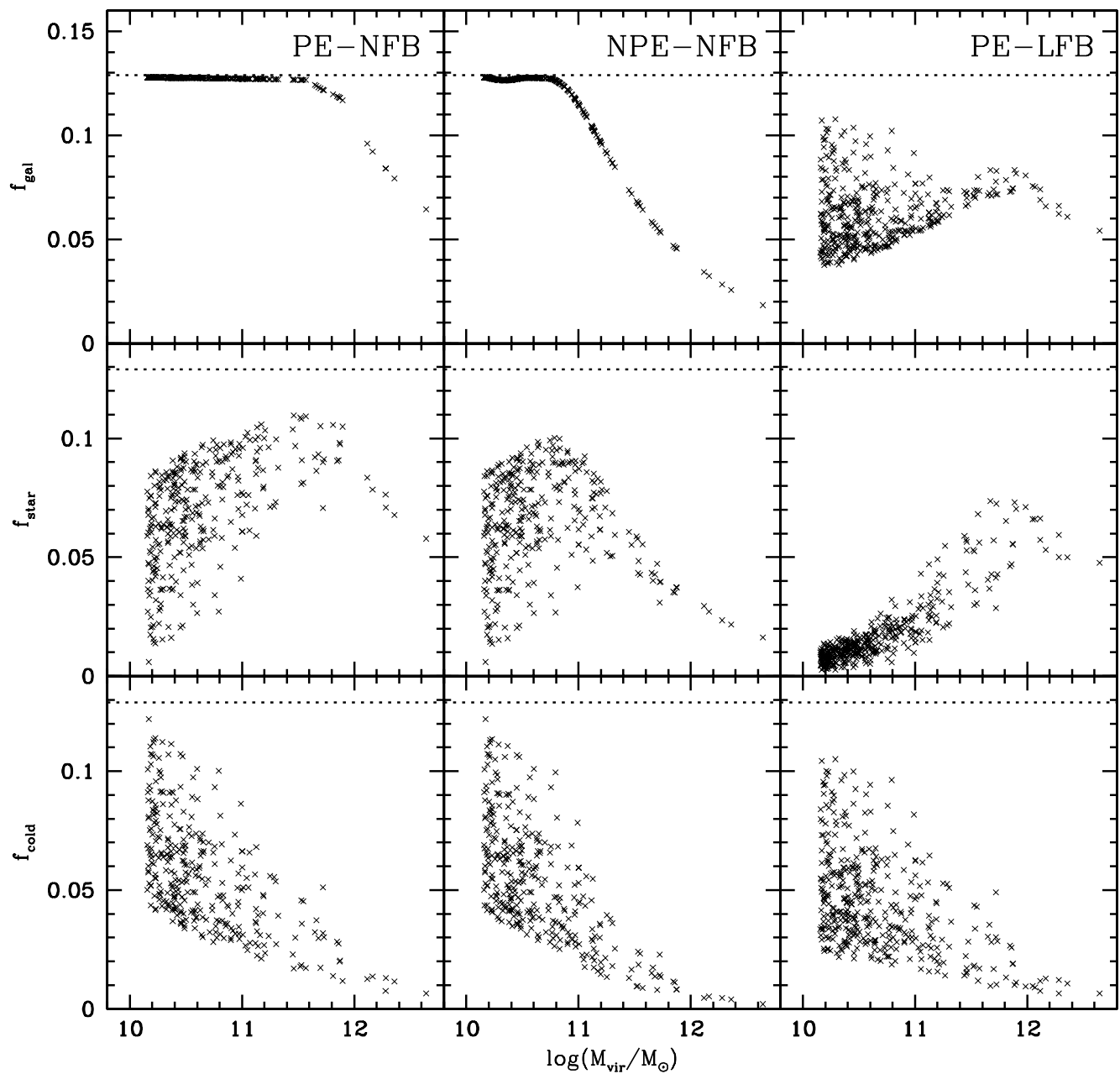,width=\hdsize}}
\caption{Upper  panels  plot the  present  day  galaxy mass  fractions
$f_{\rm  gal}$  as  function  of  virial mass  for  samples  of  three
different models:  the PreEnrichment  model with No  FeedBack (PE-NFB,
left), the No PreEnrichment,  No FeedBack model (NPE-NFB, middle), and
the  PreEnrichment Low-FeedBack Model  (PE-LFB, right).   These models
only differ  in the  feedback efficiency, the  metallicity of  the hot
gas, and  the stellar  yield, as indicated  in Table~\ref{tab:models}.
The lower two rows  of panels show how the galaxy mass  is split up in
stars  ($f_{\rm star}$)  and cold  gas ($f_{\rm  cold}$).   The dotted
lines correspond to the  universal baryon fraction $f_{\rm bar}$ which
indicates the  maximum possible  value of $f_{\rm  gal}$.  In  the two
models without  feedback (PE-NFB  and NPE-NFB), $f_{\rm  gal}$ depends
only  on  the fraction  of  baryons that  can  cool  and is  therefore
independent  of $\lambda$.   When feedback  is included  (as  in Model
PE-LFB),  $f_{\rm  gal}$ also  depends  on  the  halo spin  parameter,
resulting in  a large  amount of  scatter at the  low mass  end (where
feedback is  most effective).  Note also that  feedback impacts mainly
on $f_{\rm star}$, not on $f_{\rm cold}$.}
\label{fig:massfrac}
\end{figure*}

\section{The baryonic mass fractions of disc galaxies}
\label{sec:barfrac}

In  this  paper  we  consider  six  models that  only  differ  in  the
metallicity of  the hot gas,  $Z_{\rm hot}$, the  feedback efficiency,
$\varepsilon_{\rm  fb}$, and  the stellar  yield $y$.   The parameters
used are listed in Table~\ref{tab:models}.  All other model parameters
are     kept     fixed    at     their     fiducial    values     (see
Section~\ref{sec:models}).  Our  prime focus in  this paper will  be a
comparison  between  the  first  three models.  The  PreEnrichment  No
Feedback model (PE-NFB) with $Z_{\rm  hot} = 0.3 Z_{\odot}$, a typical
value for  the hot gas  in clusters (Mushotzsky \&  Loewenstein 1997),
and $\varepsilon_{\rm  fb}=0$, the No PreEnrichment  No Feedback model
(NPE-NFB) with  $Z_{\rm hot} = 0.0$ and  $\varepsilon_{\rm fb}=0$, and
the PreEnrichment Low-FeedBack model  (PE-LFB) with $Z_{\rm hot} = 0.3
Z_{\odot}$ and  $\varepsilon_{\rm fb}=0.02$ (i.e., two  percent of the
SN energy is converted to kinetic energy).  In each model the yield is
tuned so that the cold gas  in the fiducial MW-like model discussed in
Section~\ref{sec:fiducial} has solar  metallicity at the solar radius.
For each model we construct  samples of $400$ model galaxies.  Present
day  virial masses  are  drawn from  the  Press-Schechter (1974)  mass
function with  $10^{10} h^{-1} \Msun \leq M_{\rm  vir}(0) \leq 10^{13}
h^{-1} \Msun$.  This corresponds to haloes with circular velocities in
the range $31 \kms \leq V_{\rm  vir} \leq 312 \kms$, roughly the range
expected for galaxies.  Spin  parameters are drawn from the log-normal
distribution of equation~(\ref{spindistr}),  and for each model galaxy
we use  the Universal MAH  of equation~(\ref{unimah}).  We  are mainly
interested  in spiral  galaxies (of  type Sa  and  later).  Therefore,
galaxies  that are too  bulge-dominated are  removed from  the sample.
Following Simien  \& de Vaucouleurs  (1986) we use the  criterion that
the total  galaxy has to be  at least $0.98$  magnitudes brighter than
the bulge in the $B$-band.

\subsection{Theoretical Predictions}
\label{sec:predictions}

In the  upper panels of Figure~\ref{fig:massfrac} we  plot the present
day galaxy mass  fractions $f_{\rm gal} = M_{\rm  gal}/M_{\rm vir}$ as
function  of $M_{\rm  vir}$ (for  models PE-NFB,  NPE-NFB,  and PE-LFB
only). Here $M_{\rm gal}$ is defined as the mass of the galaxy (disc +
bulge), which includes all the  baryons inside the virial radius minus
the  hot halo  gas  and the  gas that  is  expelled from  the disc  in
SN-driven outflows.   In the PE-NFB Model (left  panels) $f_{\rm gal}$
is virtually identical to  the universal baryon fraction $f_{\rm bar}$
(indicated  by the horizontal  dotted line)  for systems  with $M_{\rm
vir}(0) \lta 5 \times 10^{11} h^{-1} \Msun$.  For more massive systems
the  cooling is less  efficient and  $f_{\rm gal}$  decreases strongly
with increasing  virial mass.  This  is even more pronounced  in model
NPE-NFB (middle  panels) where cooling  is less efficient  and $f_{\rm
gal}$ already drops below $f_{\rm  bar}$ at $M_{\rm vir} \sim 5 \times
10^{10} h^{-1}  \Msun$.  For zero-metallicity gas the  cooling time is
so long that for massive  galaxies with $M_{\rm vir}(0) \simeq 10^{13}
h^{-1}  \Msun$ only about  15 percent  of all  the baryons  inside the
virial  radius have  had sufficient  time  to cool.   In model  PE-LFB
(right  panels),  $f_{\rm gal}  \ll  f_{\rm  bar}$  for the  low  mass
systems, but with a large amount  of scatter.  This is a reflection of
the  scatter  in  halo  spin  parameters: systems  with  less  angular
momentum produce  discs with higher surface  densities, therefore have
higher   star  formation   rates,  which   induce  a   more  efficient
feedback. At the high mass end  $f_{\rm gal}$ is fairly similar to the
PE-NFB model  without feedback.  This owes  to the fact  that the mass
ejection  efficiency scales inversely  with the  square of  the escape
velocity  (see   equation~[\ref{mass_eject}]),  making  feedback  less
efficient in more massive systems.

The panels  in the middle  and lower rows  plot the mass  fractions of
stars (disc plus bulge), $f_{\rm star}$, and cold gas, $f_{\rm cold}$.
Note how $f_{\rm cold}$ depends  only weakly on both $Z_{\rm hot}$ and
$\varepsilon_{\rm   fb}$:  changes   in  the   cooling   and  feedback
efficiencies mainly influence the  stellar mass fractions, not the gas
mass fractions.
\begin{figure}
\centerline{\psfig{figure=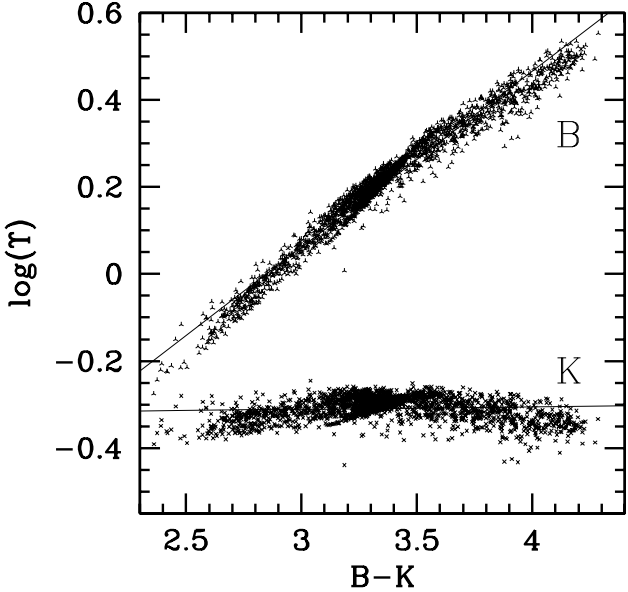,width=\hssize}}
\caption{The logarithm of the  stellar mass-to-light ratios in the $B$
(tripods) and $K$ (crosses) bands  as functions of the $B-K$ color for
all galaxies in all  six models listed in Table~\ref{tab:models}. Note
the  narrow, almost  linear, relation,  which is  fitted by  the solid
lines  (equations~[\ref{mtolB}]  and~[\ref{mtolK}]).   This  indicates
that photometric  color can  be used as  an accurate indicator  of the
stellar mass-to-light ratio (see also Bell \& de Jong 2001).}
\label{fig:mtol}
\end{figure}

As shown above $f_{\rm gal}(M_{\rm vir})$ depends strongly on both the
cooling and  feedback efficiencies. Therefore,  if one could  obtain a
measure of  $f_{\rm gal}(M_{\rm vir})$ observationally  it would allow
us to constrain the poorly understood physics of cooling and feedback.
This requires  one to be able  to infer both the  baryonic galaxy mass
$M_{\rm  gal}$ as well  as the  total virial  mass $M_{\rm  vir}$ from
observations of the luminous (and gaseous) components.  We now use our
models  to  investigate  which  observables  are best  suited  as  the
appropriate mass indicators.

\subsection{Determining the baryonic mass of disc galaxies}
\label{sec:barmass}

For  the total  baryonic  galaxy mass  one  can write  $M_{\rm gal}  =
\Upsilon_k  L_k +  M_{\rm  gas}$.   Here $L_k$  is  the luminosity  in
photometric  passband $k$, $\Upsilon_k$  is the  corresponding stellar
mass-to-light ratio,  and $M_{\rm gas}$  is the total gaseous  mass of
the galaxy.   Note that it  is apparent from  the lower two  panels of
Figure~\ref{fig:massfrac} that one can  not ignore the gas mass, which
can  easily exceed  the  total stellar  mass,  especially in  low-mass
systems.  However, measuring $M_{\rm  gas}$ is complicated by the fact
that the gas component of  (disc) galaxies is a multi-phase component,
consisting of atomic, molecular, and warm gas. Since our models do not
consider such a multi-phase medium, we can not make useful predictions
for  the  accuracy  with  which   one  can  hope  to  measure  $M_{\rm
gas}$. Therefore, we  simply assume for the moment  that $M_{\rm gas}$
can be obtained accurately  from HI observations (with the appropriate
scaling to take account of Helium), and we warn the reader that we are
thus underestimating the  uncertainties related to determining $M_{\rm
gal}$.

An  important source  of error  in  determining $M_{\rm  gal}$ is  the
unknown stellar  mass-to-light ratio $\Upsilon$. However,  in a recent
paper Bell \&  de Jong (2001), have shown  that $\Upsilon$ is strongly
correlated with  color.  In particular, they showed  that the relation
between $\Upsilon$ and color depends only weakly on the star formation
history and  on dust extinction  effects.  The main  uncertainties are
related to the unknown IMF: although the wrong IMF causes a zero-point
offset,  it  conserves  the  slope  of the  relation.   These  results
therefore suggest that using multi-color photometry one should be able
to   obtain  fairly   accurate   estimates  of   the  {\it   relative}
stellar mass-to-light ratios of disc galaxies.

In  Figure~\ref{fig:mtol}  we   plot  $\Upsilon_B$  and  $\Upsilon_K$,
averaged over the entire disc plus bulge, as function of $B-K$ for all
galaxies in all six models listed in Table~\ref{tab:models}. As can be
seen, our models, which use  the Bruzual \& Charlot stellar population
models  with a  Scalo IMF  and the  IRA for  chemical  evolution, also
reveal a  narrow correlation  between stellar mass-to-light  ratio and
$B-K$ color,  in excellent  agreement with the  results of Bell  \& de
Jong. Linear fits (indicated by thin lines) yield
\begin{equation}
\label{mtolB}
{\rm log}(\Upsilon_B) = 0.405 (B-K) - 1.155
\end{equation}
and
\begin{equation}
\label{mtolK}
{\rm log}(\Upsilon_K) = 0.005 (B-K) - 0.327
\end{equation}
We  have  also  computed   the  colors  and  mass-to-light  ratios  of
individual  radial bins  in  a  given model  galaxy,  rather than  the
parameters  averaged over  the entire  model galaxies,  and  find that
these results overlap with those plotted in Figure~\ref{fig:mtol}.  We
therefore conclude  that, modulo the  uncertainty in the  stellar IMF,
multi-color photometry  of disc galaxies  should in principle  allow a
fairly  accurate  determination of  the  total  stellar  mass of  disc
galaxies.
\begin{figure*}
\centerline{\psfig{figure=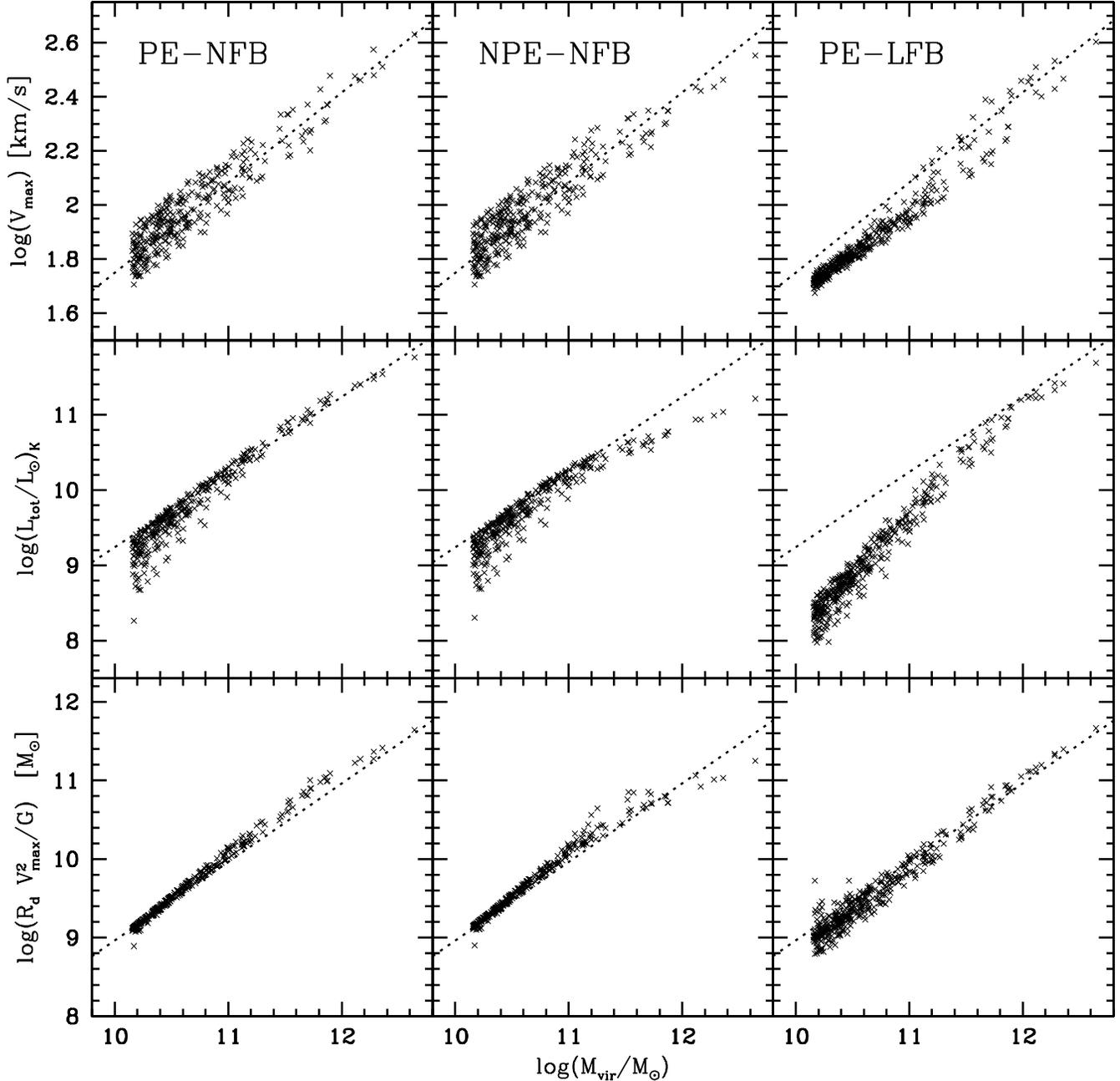,width=\hdsize}}
\caption{The relation  between various virial mass  estimators and the
actual    virial   mass    for   the    three   models    plotted   in
Figure~\ref{fig:massfrac}.   In  the  upper panels  ${\rm  log}(V_{\rm
max})$ is  plotted versus ${\rm log}(M_{\rm vir})$.   The dotted lines
correspond  to  $M_{\rm vir}  \propto  V^{1/3}_{\rm  max}$,  and is  a
reasonable description of the  average relation.  However, the scatter
is large, and the zero-point  depends on the actual model, which makes
$V_{\rm max}$ unsuitable  as virial mass indicator. The  same goes for
the  $K$-band  luminosity, which  is  plotted  in  the middle  row  of
panels. Here the dotted lines correspond to $L_K \propto M_{\rm vir}$,
which only  yields a reasonable description for  the brighter galaxies
in models with  $Z_{\rm hot} = Z_{\odot}/3$. For  fainter galaxies the
scatter  is again  large.  Furthermore  the slope  of  the $L_K(V_{\rm
vir})$ relation  depends strongly on  the feedback efficiency.  In the
lower panels we plot the virial mass estimator $R_d V^2_{\rm max} / G$
as  function of  $M_{\rm vir}$.  Here the  dotted lines  correspond to
equation~(\ref{mvir}), which provides  a reasonable description of the
models, independent of the cooling and/or feedback efficiencies.}
\label{fig:masses}
\end{figure*}

\subsection{Determining the virial mass of disc galaxies}
\label{sec:virmass}

For  virialized  systems $M_{\rm  vir}  \propto  V_{\rm vir}^3$,  with
$V_{\rm  vir}$  the  halo  circular  velocity at  the  virial  radius.
Unfortunately,  one cannot measure  the circular  velocity out  at the
virial radius.  Instead, one  can only measure the rotation velocities
of  the  stars  and  gas  in   the  disc.   In  the  upper  panels  of
Figure~\ref{fig:masses} we plot $V_{\rm  max}$, defined as the maximum
rotation velocity  inside the  radial extent probed  by the  cold gas,
versus $M_{\rm vir}$. Results correspond to $z=0$, and are plotted for
all three models.  The dotted lines correspond to $V_{\rm max} \propto
M_{\rm  vir}^{1/3}$ and  are  plotted for  comparison (with  arbitrary
zero-point).  Although the average  relation between $V_{\rm max}$ and
$M_{\rm  vir}$  nicely  follows   this  theoretical  scaling,  in  the
no-feedback models PE-NFB  and NPE-NFB a given value  of $V_{\rm max}$
has a  corresponding scatter in $M_{\rm  vir}$ of a  factor six.  This
large amount of scatter owes entirely to the scatter in $\lambda$: the
halo  angular momentum  sets the  concentration of  the  baryonic mass
component after cooling, which, together with the effects of adiabatic
contraction, causes a  large spread in $V_{\rm max}$  at given $M_{\rm
vir}$ (cf.  the lower panels of Figure~\ref{fig:sdall}).  Furthermore,
although  the  scatter in  model  PE-LFB  is  significantly less,  the
zero-point of  the $M_{\rm vir}(V_{\rm max})$ relation  is offset with
respect to the models without  feedback. We thus conclude that $V_{\rm
max}$ can  not be  used as  a reliable indicator  of the  total virial
mass.
\begin{figure*}
\centerline{\psfig{figure=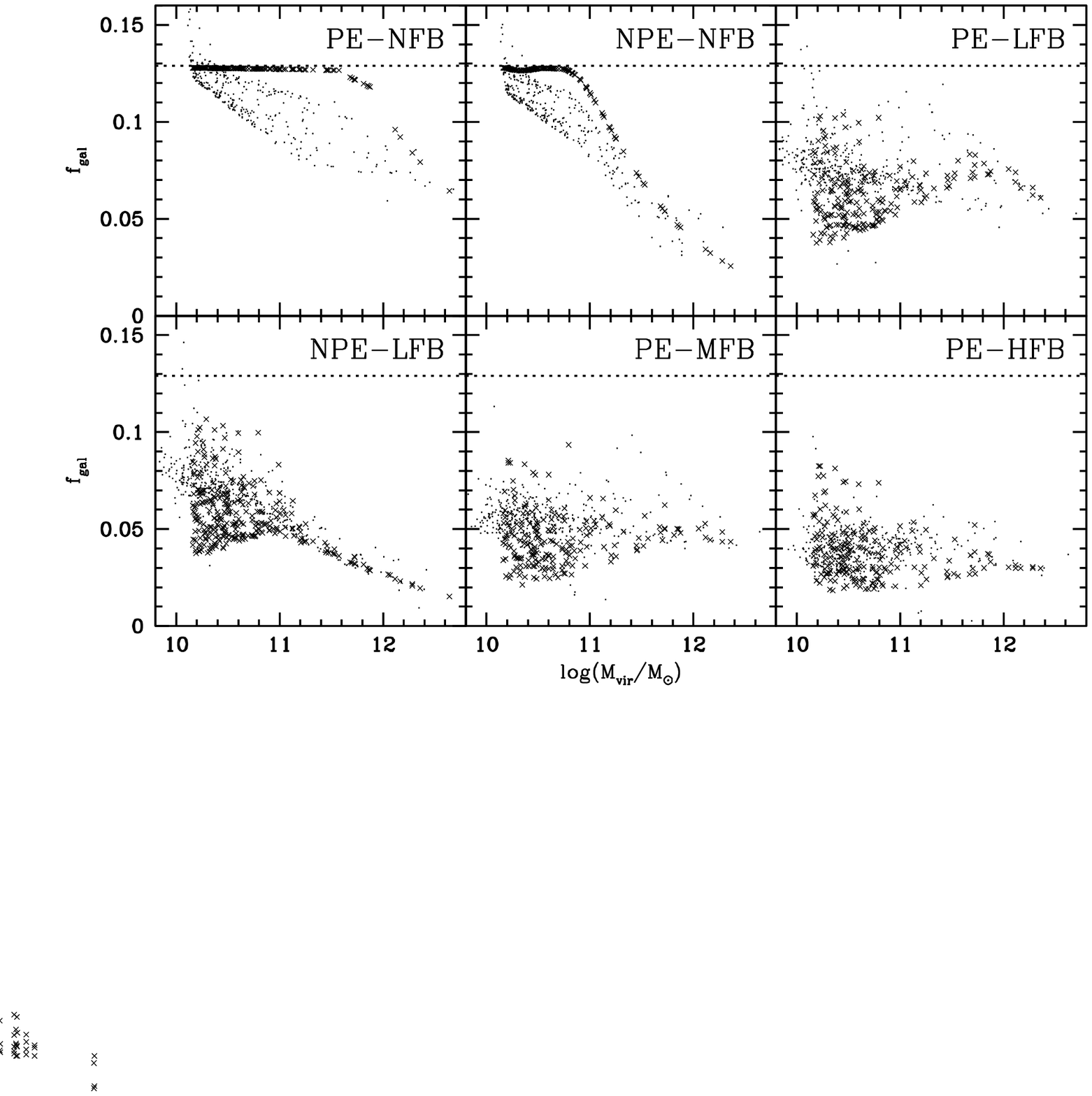,width=\hdsize,clip=}}
\caption{A comparison  of the true galaxy mass  fraction $f_{\rm gal}$
as function  of the  true virial mass  (crosses) with the  same values
estimated  from  the observables  extracted  from  the models  (dots).
Results  are plotted  for  all  six models  discussed  in this  paper.
Dotted  lines  correspond  to  the universal  baryonic  mass  fraction
$f_{\rm bar}$.  In the  models with feedback (PE-LFB, NPE-LFB, PE-MFB,
and PE-HFB) the  dots occupy the same parameter  space as the crosses,
indicating  that   the  observables  allow  one   to  recover  $f_{\rm
gal}(M_{\rm  vir})$, at  least in  a  statistical sense.   In the  two
models  without  feedback (PE-NFB  and  NPE-NFB)  the  scatter in  the
recovered  $f_{\rm  gal}(M_{\rm vir})$  is  much  larger  than in  the
intrinsic $f_{\rm  gal}(M_{\rm vir})$, but the former  still allows to
discriminate  between the  two  models; in  particular, the  estimated
galaxy  mass  fractions  nicely  avoid  the  upper  right  regions  of
parameter   space   which   contain   information   on   the   cooling
efficiencies.}
\label{fig:try}
\end{figure*}

Next we investigate how well the total $K$-band luminosity can be used
as indicator  of total virial mass  (panels in middle  row).  In model
PE-NFB  the luminous galaxies  nicely follow  a relation  $L_K \propto
M_{\rm vir}$ with very little  scatter. However, for the less luminous
systems this narrow relation breaks down.  This is a reflection of the
fact that less  massive systems have a much larger  spread in gas mass
fractions (cf.  lower  panels of Figure~\ref{fig:massfrac}).  In model
NPE-NFB  one can  see a  clear curvature  at the  high-luminosity end,
reflecting the  cooling inefficiencies.  In  model PE-LFB there  is no
longer a  linear relation between $L_K$  and $M_{\rm vir}$  due to the
higher  feedback efficiencies in  lower mass  systems. Based  on these
results  we  thus  conclude  that  total luminosity  is  also  a  poor
indicator of total virial mass.

Finally, in the lower row of panels in Figure~\ref{fig:masses} we plot
the mass measure  $R_d V_{\rm max}^2 / G$  versus $M_{\rm vir}$.  Here
$R_d$ is the disc scale  length in the $I$-band, obtained from fitting
an exponential to the  $I$-band surface brightness distribution of the
disc. The dotted lines correspond to the fitting relation
\begin{equation}
\label{mvir}
M_{\rm vir} = 2.54 \times 10^{10} \Msun \left({R_d \over \kpc}\right)
\left({V_{\rm max} \over 100 \kms}\right)^2.
\end{equation}
The zero-point of this relation  is determined by fitting to all model
galaxies of all six models  simultaneously, and is not necessarily the
best fit for  one particular model.  The model  galaxies nicely follow
this  relation,  with  an  rms  scatter  between  20  and  50  percent
(depending on the amount of feedback).  The fraction of model galaxies
for which equation~(\ref{mvir}) yields  an estimate of the true virial
mass that is off  by more than a factor two is  smaller than $8$ ($2$)
percent in models with (without) feedback.

It is remarkable  that the zero-point for models  with feedback is the
same  as for  models  without feedback.   Apparently,  when matter  is
ejected it  reduces $V_{\rm max}$ but  at the same  time increases the
disc  scale  length  such  that  $R_d  V_{\rm  max}^2$  stays  roughly
constant, albeit  with somewhat  more scatter.  We  therefore conclude
that one  can use equation~(\ref{mvir}) as a  fairly reliable estimate
of the total virial mass of disc galaxies.

In summary,  contrary to naive expectations  maximum rotation velocity
and  total luminosity  are very  poor indicators  of the  total virial
mass.   First  of  all  the  slope  and  zero-points  of  the  $V_{\rm
max}(M_{\rm  vir})$ and  $L_K(M_{\rm  vir})$ relations  depend on  the
input parameters of  the model. This means that  an observer trying to
infer $M_{\rm vir}$  from either $V_{\rm max}$ or  $L_K$ needs to make
assumptions about  the efficiencies of cooling  and feedback. However,
it is exactly these efficiencies that we seek to constrain.  Secondly,
the  scatter of  both  relations can  be  so large  that  even if  the
normalization of  the relation were  known, one could still  not infer
$M_{\rm  vir}$ to  better than  an order  of magnitude.   A  much more
reliable mass indicator  is $R_d V_{\rm max}^2 /  G$, which can fairly
easily be  obtained from observations. Although the  error in inferred
virial mass for each individual  galaxy can still exceed a factor two,
especially when feedback is important, the {\it average} ratio between
$M_{\rm  vir}$ and  $R_d V_{\rm  max}^2 /  G$, averaged  over  a large
ensemble  of galaxies,  is  independent of  the  cooling and  feedback
efficiencies.
\begin{figure*}
\centerline{\psfig{figure=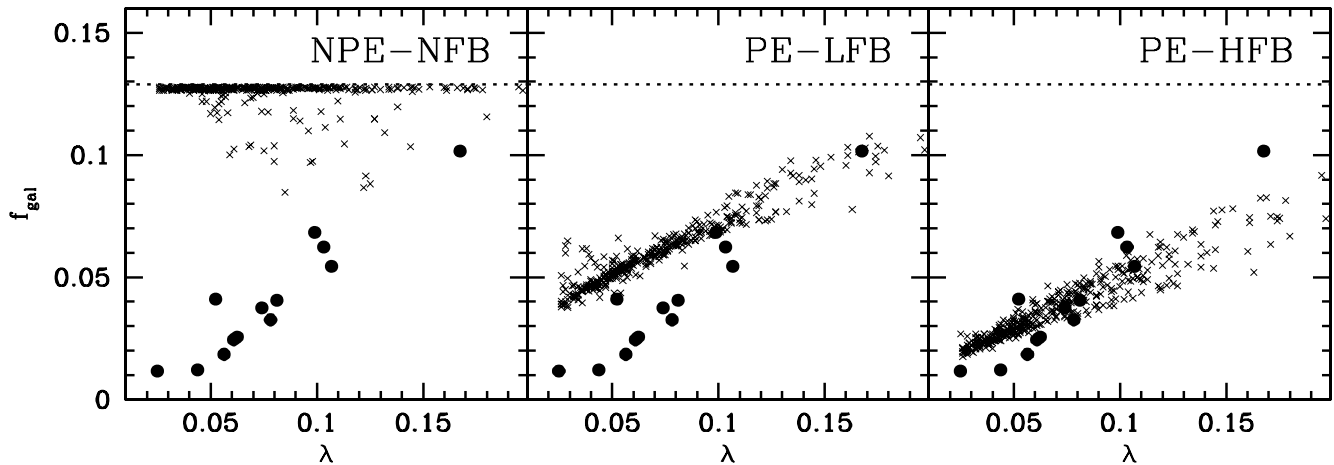,width=\hdsize}}
\caption{The relation  between halo  spin parameter $\lambda$  and the
galaxy  mass fraction  $f_{\rm gal}$  in models  NPE-NFB,  PE-LFB, and
PE-HFB (crosses).  The thick  solid dots correspond  to data  on dwarf
galaxy rotation curves  analyzed by van den Bosch,  Burkert \& Swaters
(2001).  These data reveal  a narrow correlation between $f_{\rm gal}$
and  $\lambda$, which  is  reasonably well  reproduced  by the  PE-HFB
model.   The model  without  feedback (NPE-NFB),  on  the other  hand,
predicts  a  clearly   different  relation  between$f_{\rm  gal}$  and
$\lambda$.   Note that here  low values  of $f_{\rm  gal}$ are  due to
inefficient cooling rather than efficient feedback. This suggests that
the low baryonic mass fractions  of dwarf galaxies are due to feedback
rather  than inefficient  cooling. See  the text  for a  more detailed
discussion.}
\label{fig:lam}
\end{figure*}

We  now have all  the tools  in place  to see  whether we  can recover
$f_{\rm gal}(M_{\rm vir})$ from  the observables. Using $M_{\rm gal} =
M_{\rm   cold}  +   \Upsilon_B   L_B$  with   $\Upsilon_B$  given   by
equation~(\ref{mtolB}),  and using  equation~(\ref{mvir})  to estimate
$M_{\rm     vir}$    we    obtain     the    results     plotted    in
Figure~\ref{fig:try}. This time we plot the results for all six models
listed  in  Table~\ref{tab:models}.  Crosses  correspond  to the  true
values of $f_{\rm  gal}$ and $M_{\rm vir}$, while  the dots correspond
to the  values inferred from  the observables as indicated  above.  In
both models without  feedback (PE-NFB and NPE-NFB) the  scatter in the
recovered $f_{\rm gal}(M_{\rm vir})$, which owes entirely to errors in
the  estimate of  $M_{\rm  vir}$, is  much  larger than  for the  true
values. Yet, the two  models are sufficiently different to distinguish
between  them, and  the estimates  of $f_{\rm  gal}$ nicely  avoid the
regions in the upper right corner which contains the information about
the cooling efficiencies.  In  the models with feedback, the recovered
values occupy roughly the same area of the $f_{\rm gal} - M_{\rm vir}$
plane as the intrinsic values.  Although the one-to-one correspondence
for individual model galaxies may be poor, statistically the method to
recover  $f_{\rm  gal}(M_{\rm vir})$  explored  here works  remarkably
well.

It   is   important   to   realize   that   the   results   shown   in
Figure~\ref{fig:try}  correspond  to the  idealized  case. In  reality
there  will  be additional  scatter  from  observational errors,  from
errors in estimates of the cold  gas mass (which we have taken here to
be known exactly), and from  the unknown normalization of the relation
between  color and  mass-to-light ratio  due to  uncertainties  in the
IMF. Nevertheless,  despite this  pessimistic outlook, there  is still
hope that the aforementioned method might at least be able to put some
limits  on  the  feedback   efficiencies,  as  the  resulting  $f_{\rm
gal}(M_{\rm  vir})$  relation  of  the models  with  feedback  clearly
separates out from the two  models without feedback.  Although it will
be extremely difficult to distinguish between medium and high feedback
(i.e., models PE-MFB and PE-HFB), this is not a problem of the method,
but  owes  mainly  to  the   fact  that  even  the  intrinsic  $f_{\rm
gal}(M_{\rm vir})$  relations of these two models  are fairly similar.
This  in turn  reflects a  saturation of  the effective  efficiency of
feedback in expelling mass: stronger mass ejection reduces the mass of
new stars formed, which in turn reduces the mass ejection efficiency.

\subsection{Correlations with halo spin parameter}
\label{sec:flambda}

The  two main parameters  that determine  the characteristics  of disc
galaxies are  the total mass  and the halo  spin parameter. So  far we
have focussed on $f_{\rm  gal}(M_{\rm vir})$, and we discussed various
indicators of  $M_{\rm vir}$.  Here  we turn our attention  to $f_{\rm
gal}(\lambda)$.

In  Figure~\ref{fig:lam} crosses  plot  the galaxy  mass fractions  as
function of  the halo  spin parameter for  models NPE-NFB,  PE-LFB and
PE-HFB.  Only model  galaxies with $50 \kms \leq  V_{\rm max} \leq 150
\kms$ are plotted.  Models with  and without feedback reveal a clearly
different $f_{\rm gal}(\lambda)$. In the latter, low values of $f_{\rm
gal}$ are due to inefficient  cooling, which is unrelated to the value
of  the spin  parameter.  However,  in cases  with feedback,  a narrow
correlation is  evident whereby galaxies in  low-$\lambda$ haloes have
lower  galaxy   mass  fractions.   This  was  already   eluded  to  in
Section~\ref{sec:predictions}, and is a result of the fact that haloes
with   less  angular  momentum   produce  higher   surface  brightness
discs. Since the star  formation efficiency is correlated with surface
density  (equation~[\ref{Schmidt_law}])  and  the  amount  of  ejected
material   is   proportional   to   the   amount   of   stars   formed
(equation~[\ref{mass_eject}]), systems with  less angular momentum are
more efficient in ejecting mass. Decreasing or increasing the feedback
efficiency  $\varepsilon_{\rm  fb}$ moves  the  crosses  up and  down,
respectively,  but  leaves the  slope  of  the $f_{\rm  gal}(\lambda)$
relation  intact.   Therefore,  if   we  could  somehow  measure  both
$\lambda$ and $f_{\rm gal}$ we could use that to constrain the cooling
and feedback efficiencies.

In principle, both the spin parameter  of the baryons that make up the
disc and  the galaxy mass fraction  can be determined  from a detailed
rotation curve analysis. In a  recent study, van den Bosch, Burkert \&
Swaters  (2001) applied  such  method to  a  sample of  low mass  disc
galaxies  (also with $50  \kms \leq  V_{\rm max}  \leq 150  \kms$) for
which  accurate  HI  rotation   curves  and  $R$-band  photometry  are
available.  Fitting  mass models to the observed  rotation curves, van
den  Bosch  \etal (2001)  obtained  estimates  of  both $\lambda$  and
$f_{\rm  gal}$  which are  plotted  in  Figure~\ref{fig:lam} as  solid
dots. Remarkably  enough, these data points reveal  a similarly narrow
relation, albeit  with a  somewhat steeper slope,  as our  models with
feedback. Taking these data point at face value thus suggests that the
low  baryonic  mass  fractions  in  dwarf  galaxies  are  not  due  to
inefficient cooling, but to efficient feedback.

However,  it is  important to  realize  that the  values of  $\lambda$
plotted in  Figure~\ref{fig:lam} correspond to the  spin parameters of
the dark matter  haloes in the case  of the model, but to  that of the
disc material (cold  gas plus stars) in the case of  the data.  In the
standard  picture  of disc  formation  (adopted  in  our models),  the
assumption  is  made  that  baryons conserve  their  specific  angular
momentum. Therefore, if all baryons end  up in the disc, halo and disc
should have the  same spin parameter. However, since  $f_{\rm gal} \ll
f_{\rm bar}$ this  clearly is not the case, implying  that one can not
simply  compare the  spin parameters  of halo  and disc.   In  fact, a
straightforward  interpretation of  Figure~\ref{fig:lam} for  the data
seems to  imply that discs  form out of  only a small fraction  of the
available  baryons, but  yet  manage  to draw  most  of the  available
angular momentum (see discussions in Navarro \& Steinmetz 2000 and van
den Bosch \etal 2001).  However, the disc spin parameter is determined
by integrating  the ratio $M_{\rm  disc}(r)/M_{\rm disc}(r_{\rm max})$
over the  entire disc  (see van den  Bosch \etal 2001).   Here $M_{\rm
disc}(r)$ is the  total disc mass inside radius  $r$ and $r_{\rm max}$
is the maximum extent of the disc.  Therefore, $\lambda_{\rm disc}$ is
still  identical to  $\lambda_{\rm halo}$  if,  and only  if, at  each
radius in the  disc feedback has expelled an  identical {\it fraction}
of disc  material. Surprisingly, as  shown in paper~1,  our simplistic
feedback model  establishes just that.  Therefore, if  feedback in the
real  Universe  accomplishes a  similar  radial  dependence, a  direct
comparison of model with data as in Figure~\ref{fig:lam} is justified.

Finally  we  emphasize that  the  errors  on  both $f_{\rm  gal}$  and
$\lambda$ inferred from data can  be quiet significant (though hard to
quantify).  Both $\lambda$ and $f_{\rm gal}$ are derived directly from
the model fits to the  observed rotation curves.  However, as shown in
van den Bosch \& Swaters (2001), such mass models suffer from numerous
uncertainties and degeneracies.  For instance, inferring $M_{\rm vir}$
from  the observed rotation  curve, requires  one to  make assumptions
regarding the density distribution of the dark matter halo (which sets
the  rotation curve  shape), and  cosmology (which  sets  the relation
between virial mass and virial radius, the latter of which is required
for the  computation of $\lambda$).   Furthermore, the scatter  in the
$f_{\rm gal}(\lambda)$  relation of the model galaxies  is also likely
to be underestimated. For instance, the  scatter in MAHs for a halo of
given present  day mass is  likely to cause  some scatter in  the halo
concentrations  and the  distributions of  specific  angular momentum,
even for haloes with the same spin parameter (e.g., Jing 2000; Gardner
2001; van den Bosch 2001;  Vitvitska \etal 2001; Wechsler \etal 2001),
both  of which  will  cause  some additional  scatter  in the  $f_{\rm
gal}(\lambda)$ relation. Since we adhere to the average, universal MAH
and to  the average halo concentration  given by the  model of Bullock
\etal (2001c), this is not  taken into account in the models. However,
simple tests indicate  that these effects are not  very important, and
furthermore, the scatter  in the model galaxies is  comparable to that
of the data. This suggests  that this technique might actually be able
to  provide meaningful  constraints on  feedback efficiencies,  and it
warrants  a more  detailed exploration  of the  $f_{\rm gal}(\lambda)$
relation for a larger sample of disc galaxies.
\begin{figure*}
\centerline{\psfig{figure=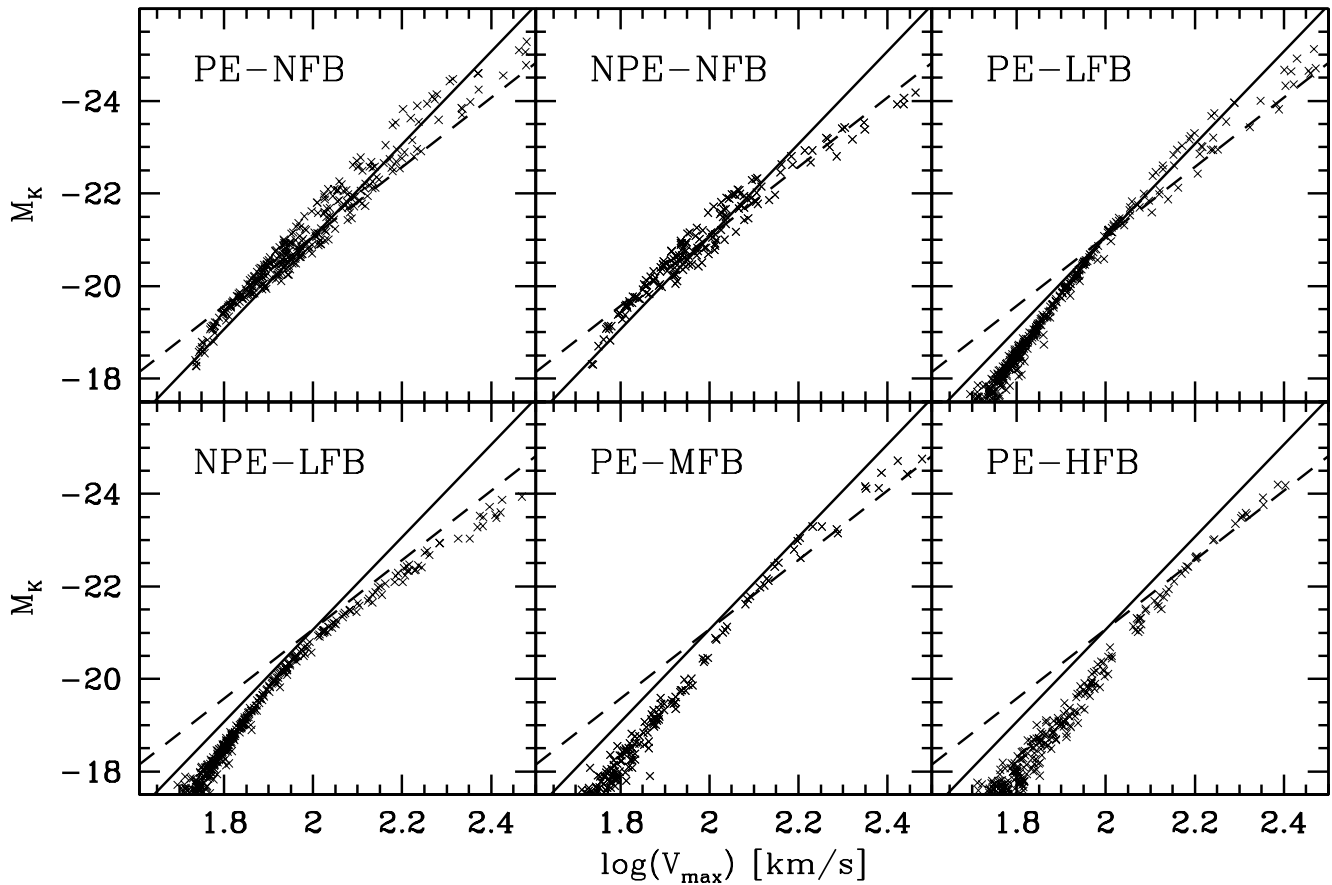,width=\hdsize}}
\caption{$K$-band Tully-Fisher relations for  the six models listed in
Table~\ref{tab:models}.   Crosses correspond  to  the model  galaxies,
while the  solid (dashed)  lines correspond to  TFRs of the  form $L_K
\propto  V^a_{\rm  max}$  with  $a=4$  ($a=3$), and  are  plotted  for
comparison. In  fact, the  solid lines correspond  to the  observed TF
relation of Verheijen  (2001), while the dashed lines  correspond to a
fiducial TF  relation with $a=3$ with the  zero-point normalization of
Verheijen (2001)  at $V_{\rm max}  = 100 \kms$.  Note how most  of the
models predict some amount of curvature in the TFR, steepening from $a
\simeq 3$ at the bright end to $a \simeq 4$ at the faint end.  See the
text for a detailed discussion.}
\label{fig:tf}
\end{figure*}

\section{The Tully-Fisher relation}
\label{sec:tf}

The  fundamental  scaling relation  of  disc  galaxies,  known as  the
Tully-Fisher  (hereafter TFR)  relation, couples  the  luminosities of
disc galaxies  to their rotation velocities. If  light scales linearly
with total galaxy mass, and  rotation velocity with total virial mass,
then the  slope, scatter  and zero-point of  the TFR tell  us directly
about the  efficiencies of  cooling and feedback.   Therefore, various
authors in  the past  have used detailed  models for the  formation of
disc galaxies to  try and understand the origin of  the TFR (e.g., Mo,
Mao \& White 1998; Avila-Reese \& Firmani 2000; Firmani \& Avila-Reese
2000;  Mo \&  Mao  2000; van  den  Bosch 2000;  Buchalter, Jimenez  \&
Kamionkowski 2001). Although there is a general consensus that the TFR
is governed by the relation between virial mass and circular velocity,
it is  still unclear what the  observed TFR actually  teaches us about
cooling and  feedback efficiencies. One  of the main reasons  for this
lack of  consensus is  the discordant use  of luminosity  and rotation
measures in TFRs.  The slope and scatter of the TFR depend strongly on
both photometric band and on whether  one uses HI line widths or other
measures  of the  galaxy's rotation  velocity (e.g.,  Tully,  Mould \&
Aaronson  1982; Pierce  \& Tully  1988; Gavazzi  1993;  Courteau 1997;
Verheijen 2001).  This emphasizes that  it is of crucial importance to
extract the  proper `observables' from the models  when comparing them
to data,  something that is  often ignored.  An additional  reason why
the interpretation of  the TFR is still heavily  debated is related to
the fact  that luminosity and rotation velocity  not necessarily scale
linearly with galaxy mass  and virial mass, respectively. For example,
as shown in Section~\ref{sec:barfrac} these naive expectations are not
fulfilled in our models. This  implies that the characteristics of the
TFR become sensitive  to the more subtle details  of the models, which
can differ substantially from one researcher to the other.
\begin{figure}
\centerline{\psfig{figure=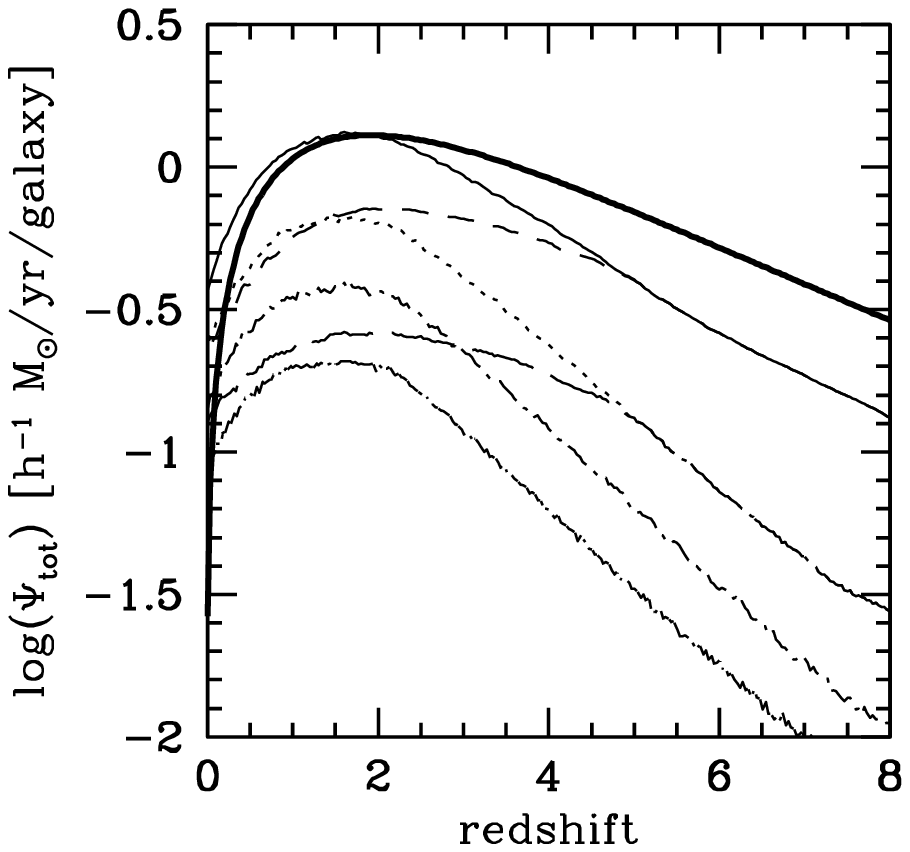,width=\hssize}}
\caption{The average star formation  rates (averaged over all galaxies
in  each sample)  as function  of  redshift.  The  various thin  lines
correspond to the  six models: PE-NFB (thin solid  line), NPE-NFB (short-dashed
line),  NPE-LFB (long-dashed  line), PE-LFB  (dotted line),  PE-MFB  (short-dash --
dotted line), and PE-HFB (long-dash -- dotted line).  The thick solid line
corresponds to  the average  {\it mass} accretion  rate, lowered  by a
factor ten to  facilitate a comparison with the  star formation rates.
Note how small changes in the feedback and cooling efficiencies impact
strongly on the star formations rates.}
\label{fig:sfh}
\end{figure}

In Figure~\ref{fig:tf}  we plot the present day  $K$-band TF relations
for  our  six  models.   We  choose  the $K$-band  since  it  is  less
susceptible  to uncertainties related  to stellar  population modeling
and  dust  extinction.  The  solid  and  dashed  lines in  each  panel
correspond to TF relations of the form
\begin{equation}
\label{tfemp}
L_k = b \left( {V_{\rm max} \over 100 \kms} \right)^{a}
\end{equation}
and are plotted for comparison.   The solid lines correspond to the TF
relation  of  the RC/FD  sample  of  Ursa  Major cluster  galaxies  of
Verheijen  (2001) for which  $a=4$ and  $b=6.19 \times  10^{9} \Lsun$.
The  dashed lines  correspond to  a  fiducial TF  relation with  $a=3$
normalized  to the  TF relation  of Verheijen  at $V_{\rm  max}  = 100
\kms$.  Although  we have  in no  way attempted to  fit our  models to
reproduce any observed TFR, it  is reassuring that our zero-points are
in reasonable agreement  with the empirical TF relation  of Ursa Major
cluster galaxies.  In the PE-NFB model  the TFR has a  slope $a \simeq
4$.  Although  the scatter in $V_{\rm  max}$ at given  $M_K$ is fairly
large, it is  significantly smaller than the scatter  in $V_{\rm max}$
at given $M_{\rm vir}$  shown in Figure~\ref{fig:masses}: more compact
discs  have relatively  higher $V_{\rm  max}$,  but at  the same  time
convert a larger  fraction of their gas into  stars, and are therefore
brighter.   Variation   in  $\lambda$  therefore   partially  scatters
galaxies  with identical  $M_{\rm  vir}$ along  the  TFR, rather  than
perpendicular to it  (see also van den Bosch  2000).  The introduction
of feedback  helps to  further suppress the  amount of scatter  in the
TFR,  as its  mass  ejection efficiency  is  strongly correlated  with
$\lambda$ (cf. Section~\ref{sec:flambda}).

More  importantly,  introducing   feedback  or  reducing  the  cooling
efficiencies both introduce  a curvature in the TFR;  the luminous end
of the TFR  has a slope $a  \simeq 3$, which steepens to  $a \simeq 4$
towards  lower  luminosities.   If  the  true TFR  reveals  a  similar
behavior than  (i) the observed slope  of the TFR  depends strongly on
the sample  of galaxies used (i.e.,  on the relative  number of bright
and faint  galaxies), and  (ii) it will  be impossible  to distinguish
between efficient  feedback and inefficient  cooling based on  the TFR
alone. Furthermore, the TFR contains very little information about the
actual feedback  efficiency (i.e., compare models  PE-LFB, PE-MFB, and
PE-HFB). Increasing  $\varepsilon_{\rm fb}$ lowers  both $V_{\rm max}$
and $L_K$,  such that  it mainly shifts  model galaxies along  the TFR
(see also  Mo \& Mao 2000) .   We therefore conclude that  the TFR not
necessarily provides  the most useful  constraint on theories  of disc
formation (but see Shen, Mo \& Shu 2001 for a possible way of breaking
these  various  degeneracies).    Secondly,  the  results  shown  here
indicate that  it is extremely  useful to obtain  a TFR that  spans as
large  a range  in luminosities  as possible,  for only  then  can one
conclusively infer whether  or not the TFR reveals  a curvature (i.e.,
most observed TF relations only  probe down to $V_{\rm max} \simeq 100
\kms$).

\section{Star formation rates, colors and metallicities}
\label{sec:colors}

As we  have shown  in Section~\ref{sec:lambdaMAH}, the  star formation
rates of disc  galaxies are largely governed by the  rate at which the
galaxies  accrete mass.   Therefore,  one  might hope  to  be able  to
reconstruct  the   mass  accretion  history  of   disc  galaxies  from
observations  of their star  formation rate  as function  of redshift.
However, the  efficiencies of cooling and feedback  will influence the
strength of  the coupling between  mass accretion and  star formation.
In order  to investigate how strong  these effects are  we compute the
average star formation rate (averaged  over all model galaxies in each
sample)  for  each of  the  models  listed in  Table~\ref{tab:models}.
Results   are  shown  in   Figure~\ref{fig:sfh}  (thin   lines).   For
comparison,  we also plot  (as a  thick solid  line) the  average mass
accretion rate,  which is identical  for all six models.   Although to
zeroth order  the overall shapes  of the star formation  histories are
similar to that  of the mass accretion history,  the SFHs are strongly
model dependent.  First of all, the absolute values of the SFR depends
strongly  on the  feedback efficiency:  for  example, the  SFR in  the
PE-HFB  model  with  $\varepsilon_{\rm  fb}  = 0.1$  is  an  order  of
magnitude lower than for the PE-NFB model without feedback.  Secondly,
reducing the metallicity of the  hot gas impacts strongly on the SFRs:
the  `no pre-enrichment' models  NPE-NFB and  NPE-LFB both  yield much
lower   SFRs  at  low   redshift  than   models  PE-NFB   and  PE-LFB,
respectively, in which the hot gas is pre-enriched to one-third Solar.
Finally, whereas the mass accretion rate drops rapidly for $z \lta 1$,
the  decrease in  the  SFR is  much  less dramatic.   This reflects  a
weakness of  the coupling between  mass accretion and  star formation.
Even after mass accretion is completely quenched, cold gas in the disc
can  still continue  to form  stars,  and hot  gas in  the halo  still
continues to cool  and provide new material for  star formation to the
disc.   These results  clearly show  that the  `cosmic  star formation
history'  (Lilly \etal 1996;  Madau, Pozzetti  \& Dickinson  1998) not
only  depends  on  cosmology  (which  sets  the  halo  mass  accretion
histories), but  also on details  related to the cooling  and feedback
efficiencies, making  an interpretation  of the cosmic  star formation
history highly degenerate.

\begin{figure*}
\centerline{\psfig{figure=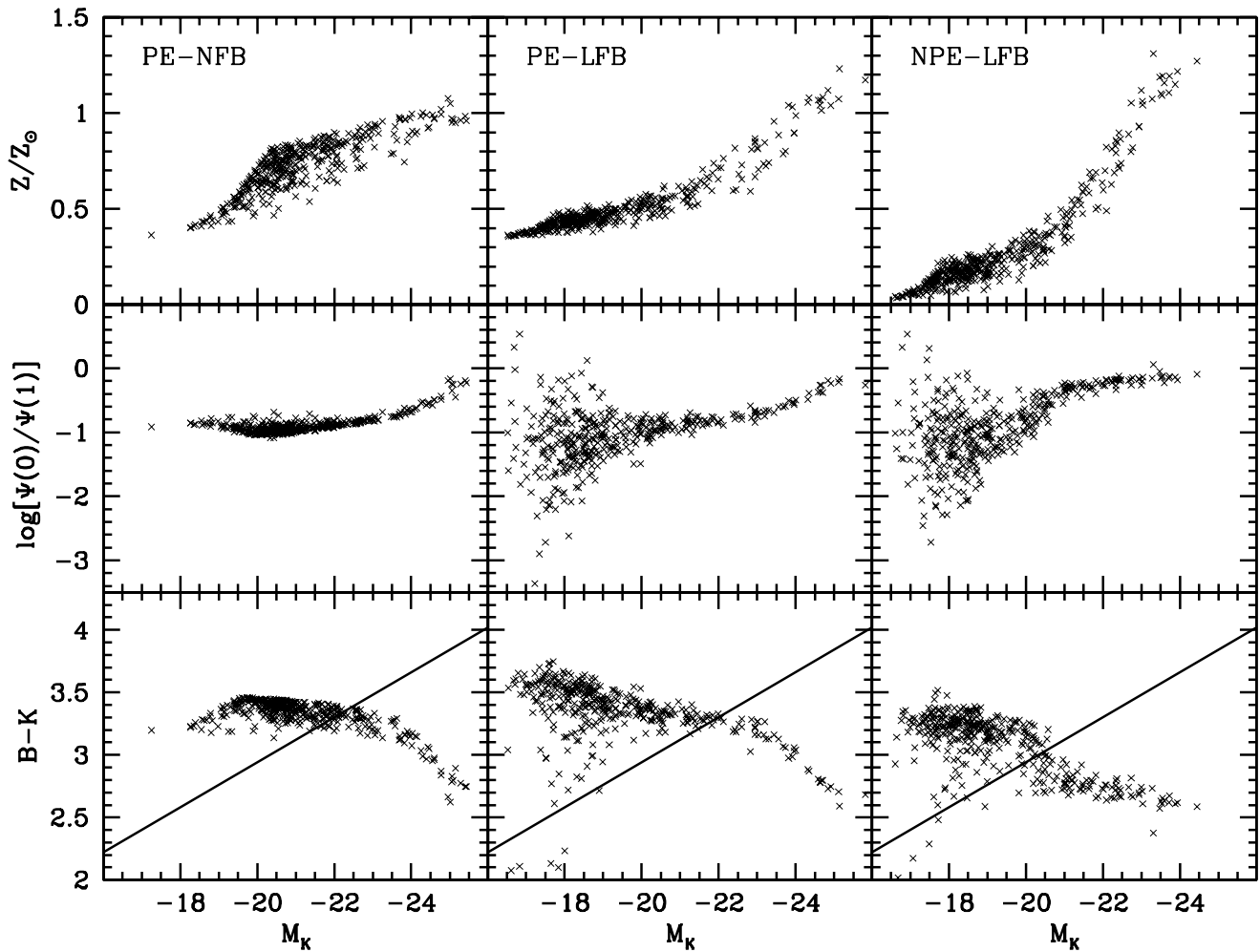,width=\hdsize}}
\caption{The upper panels plot the present day metallicity of the cold
gas (in units  of solar metallicity) as function  of absolute $K$-band
magnitude for  three representative  models. In all  cases there  is a
clear   correlation   in   which   brighter   galaxies   have   higher
metallicities.  In the models  with feedback (PE-LFB and NPE-LFB) this
owes mainly to the fact  that more massive galaxies are more efficient
at  retaining metals  (i.e.,  less ejection).   In  the model  without
feedback, this reflects the  higher efficiency of more massive systems
to turn their cold gas into stars (i.e., more enrichment).  The panels
in the  middle row plot the  ratio of the  present day SFR to  that at
$z=1$.  In the PE-NFB model  there is a narrow correlation with $M_K$,
reflecting the mass dependence of  the MAHs of dark matter haloes.  In
the models with feedback  star formation is semi-stochastic, causing a
large spread  in $\Psi(0)/\Psi(1)$ for faint  galaxies.  Bottom panels
plot the present day $B-K$  color versus $K$-band magnitude. The solid
lines  correspond to  the observed  color-magnitude relation  (van den
Bosch \&  Dalcanton 2000).  Our models clearly  fail to  reproduce the
observed  color-magnitude  relation of  disk  galaxies.  In fact,  the
models  reveal  an  inverted  relation, with  more  luminous  galaxies
becoming bluer (see discussion in text).}
\label{fig:colors}
\end{figure*}

In addition  to the  star formation rates,  variations in  cooling and
feedback efficiencies also impact  on the chemical enrichment history.
Such effects should be observable through the colors and metallicities
of the galaxies.  In  Figure~\ref{fig:colors} we plot some results for
three   representative   models:    PE-NFB,   PE-LFB,   NPE-LFB   (see
Table~\ref{tab:models}).

The  upper  panels of  Figure~\ref{fig:colors}  plot  the present  day
metallicity of  the cold gas in  the disc as function  of the absolute
$K$-band magnitude.  All  models reveal a clear metallicity-luminosity
relation in  which brighter galaxies have  higher (gas) metallicities.
In  models with  feedback, this  owes largely  to the  fact  that more
massive systems  are more efficient  in retaining the metals.   In the
PE-NFB model  no feedback is  included, but yet the  brighter galaxies
are  more enriched.   This  is  a consequence  of  the star  formation
threshold  density, which induces  larger gas  mass fractions  in less
massive     systems     (cf.      Figure~\ref{fig:massfrac}).      The
metallicity-luminosity   relations   of   the   various   models   are
sufficiently different  that there is hope that  the observed relation
might help to  constrain models of disc formation.   In particular, in
the    NPE-LFB    model   faint    galaxies    have   extremely    low
metallicities. Such galaxies are absent in models in which the hot gas
is pre-enriched,  and metallicities  of low luminosity  dwarf galaxies
should therefore  in principle  allow to put  limits on the  amount of
pre-enrichment.  However, the details of our results depend clearly on
the IRA used  and on the value of the stellar  yield (which depends on
the IMF).  Unless a more  realistic chemical enrichment model is used,
and  we can independently  constrain the  stellar yield,  we therefore
remain skeptical that the observed metallicity-luminosity relation can
place stringent constraints.

The panels in the middle row of Figure~\ref{fig:colors} plot the ratio
$\Psi(0)/\Psi(1)$ of  the present day SFR  to that at  $z=1.0$. In the
PE-NFB model there is a  smooth trend running from $\Psi(1) \simeq 0.1
\Psi(0)$  at  the  faint  end  to $\Psi(1)  \simeq  \Psi(0)$  for  the
brightest galaxies. This is a direct reflection of the mass dependence
of the MAH: less massive  systems form earlier, and therefore reveal a
declining  star formation  rate for  $z  \lta 1$,  while more  massive
systems form later and are  still actively forming stars today. In the
two models with feedback (PE-LFB  and NPE-LFB) there is a large amount
of scatter  in $\Psi(0)/\Psi(1)$ for  faint galaxies. This  scatter is
not  related  to the  scatter  in  halo  spin parameters,  but  rather
reflects the semi-stochastic star  formation history of these systems:
after  the onset  of  star formation,  feedback  ejects a  significant
fraction of the  cold gas, lowering the surface density  of the gas to
below the star formation threshold level. Consequently, star formation
is quenched until  enough new gas has cooled that  new stars can form.
This `feedback'-loop  causes the SFRs  to fluctuate wildly  with time,
which   reflects   itself   in   a   large  amount   of   scatter   in
$\Psi(0)/\Psi(1)$.

The  luminosity dependence  of  the star  formation histories  impacts
directly on  the color-magnitude  diagrams of the  resulting galaxies,
which  are plotted  in  the lower  panels of  Figure~\ref{fig:colors}.
Since the more massive systems form their stars relatively late, their
stellar populations  are relatively younger, and thus  bluer.  This is
opposite to the observed trend, in which faint galaxies are bluer than
bright galaxies  (e.g., Gavazzi  1993; Fioc \&  Rocca-Volmerange 1999;
Schombert,  McGaugh \& Eder  2001). This  is shown  by the  solid line
which corresponds to  the color magnitude relation derived  by van den
Bosch \&  Dalcanton (2000).  These  authors compiled $B$ and  $K$ band
magnitudes from the literature for  a sample of 139 spiral galaxies of
type Sb or  later, spanning the magnitude interval  $-16 \lta M_K \lta
-26$, and corrected these for external extinction.

Introducing feedback only aggravates  the problem: the SN induced mass
ejection quenches later star  formation, thus producing faint galaxies
that are extremely  red.  Only a small fraction  of the dwarf galaxies
in  the PE-LFB  and NPE-LFB  models  have sufficiently  low values  of
$\Psi(0)/\Psi(1)$ that their $B-K$ color is as blue as observed.  Even
extremely  strong  metallicity-luminosity relations,  such  as in  the
NPE-LFB model,  can not reverse the color-magnitude  relation which is
completely  dominated by  the mass  dependence of  the  star formation
histories.   Even  though faint  galaxies  in  the  NPE-LFB model  are
significantly bluer than in the  PE-LFB model (because they have lower
metallicities),  they  are  still  redder  than  their  more  luminous
counterparts.  This  indicates a generic problem  for any hierarchical
picture of galaxy formation in which more massive systems form later.

A similar problem, but for early-type galaxies, has been identified in
semi-analytical  models  of galaxy  formation  (e.g., Kauffmann  \etal
1993;  Baugh, Cole  \&  Frenk  1996; Cole  \etal  2000). Kauffmann  \&
Charlot (1998) have shown that  this problem can be solved when strong
feedback is included in the models, inducing a strong mass-metallicity
relation.  However,  as is evident  from the NPE-LFB model,  this does
not work  for disk galaxies.   The reason is that  early-type galaxies
have much  older stellar populations than disk  galaxies. A difference
between 3 and 5 Gyrs implies a much stronger color difference than one
between  8 and  12 Gyr.   Consequently, in  disk galaxies,  which have
relatively young stellar populations, the mass-metallicity relation is
not strong  enough to invert the  color-magnitude relation. Therefore,
the  color-magnitude relation  problem  is more  severe for  late-type
galaxies than for early-type galaxies.

Part of  the problem may  be solved by  including dust in  the models.
Since  more massive systems  have higher  metallicities, it  is likely
that they contain  more dust, and should therefore  be relatively more
reddened than faint galaxies.  However,  the problem is not only that
bright galaxies are too blue  (which might be solved by including dust
extinction), but also that the  faint galaxies are too red.  We intend
to  return to  this intriguing  problem in  a future  paper.   For the
moment  we   emphasize  that   variations  in  cooling   and  feedback
efficiencies do  leave signatures in  the metallicities and  colors of
galaxies,  but  that there  is  little  hope  to use  observations  to
constrain these efficiencies before  we have a better understanding of
the observed color-magnitude relations of disc galaxies.

\section{Conclusions}
\label{sec:concl}

Currently the  largest uncertainties in galaxy  formation modeling are
related   to  the   efficiencies   of  cooling,   feedback  and   star
formation. In particular, we need to understand how these efficiencies
regulate  what  fractions  of  available baryons  are  converted  into
luminous matter,  what fractions  are ejected out  of the  dark matter
haloes by  feedback processes,  and what fractions  remain in  the hot
phase because of inefficient cooling.  These efficiencies are expected
to   depend  on   both  the   mass   and  angular   momentum  of   the
protogalaxies. Therefore, if we can somehow determine the total virial
mass  and angular  momentum  of galaxies  from  observations of  their
luminous component this allows us  to put stringent constraints on the
baryonic physics that play an  important role in the process of galaxy
formation.

In  this paper  we  used models  for  the formation  of disc  galaxies
presented by  van den Bosch (2001) to  investigate which observables,
extracted directly from those models, are best suited to constrain the
initial  conditions   (i.e.,  mass   and  angular  momentum)   of  the
protogalaxies. Rather than attempting to  fit the models to real data,
we examined how variations in the efficiencies of cooling and feedback
impact on the  models. The models assume that  dark matter haloes grow
smoothly in mass  (no mergers) with a rate  that depends on cosmology.
Inside  the virialized  haloes  gas cools  and  conserves its  angular
momentum (which  owes from cosmological  torques), thus settling  in a
disc component. No  a priori assumption is made  regarding the density
distribution  of the  disc.   The models  take  star formation,  bulge
formation, chemical  evolution, and mass ejection due  to energy input
from SNe (feedback) into account,  and output masses (of six different
components),  luminosities (in  various  photometric bands),  rotation
velocities,  and   metallicities  as   functions  of  both   time  and
galactocentric radius.

The  two  main  parameters  that  we  varied in  this  paper  are  the
metallicity  of the  gas  prior  to cooling  (which  sets the  cooling
efficiency) and the  fraction of SN energy that is  used to drive mass
outflow (which  sets the  feedback efficiency).  These  two parameters
strongly impact on the galaxy  mass fraction $f_{\rm gal}$, defined as
the fraction  of total virial  mass that is  part of the  galaxy (disc
plus bulge), and in particular  on how $f_{\rm gal}$ depends on virial
mass and halo spin parameter. Therefore, if observations could somehow
determine  $f_{\rm gal}(M_{\rm  vir},\lambda)$ it  would  place strong
constraints on  theories of  galaxy formation.  We  therefore examined
which of the  observables extracted directly from the  models are best
suited to reconstruct $f_{\rm gal}(M_{\rm vir},\lambda)$ without prior
knowledge of the feedback and cooling efficiencies. The main challenge
is to  find a proper  mass estimator of  $M_{\rm vir}$. We  have shown
that both $V_{\rm  max}$ (the maximum of the  galaxy's rotation curve)
and total luminosity are poor indicators of $M_{\rm vir}$: the scatter
of the $V_{\rm max}(M_{\rm vir})$ and $L_K(M_{\rm vir})$ relations can
be extremely large,  while at the same time  the slope and zero-points
of the relations depend on  the actual model input parameters.  A much
more reliable  mass indicator  is $R_d V_{\rm  max}^2 / G$,  which can
fairly easily  be obtained from  observations.  Although the  error in
inferred virial  mass for  each individual galaxy  can still  exceed a
factor two,  especially when feedback  is important, we find  the {\it
average}  ratio between  $M_{\rm vir}$  and $R_d  V_{\rm max}^2  / G$,
averaged over a  large ensemble of galaxies, to  be independent of the
cooling and  feedback efficiencies. In  fact, we have shown  that with
this estimator of the total  virial mass, we can recover the intrinsic
$f_{\rm  gal}(M_{\rm vir})$ accurately  enough to  distinguish between
the various models.

Given the difficulties with  determining $M_{\rm vir}$ many studies in
the  past have  studied  the direct  relation  between luminosity  and
rotation velocities,  known as the  Tully-Fisher relation, to  try and
constrain galaxy formation. Most of the models presented in this paper
reveal  TFRs that  are curved:  they change  from $L  \propto V^3_{\rm
max}$ at  the bright  end to  $L \propto V^4_{\rm  max}$ at  the faint
end. In order to detect such curvature observationally it is essential
that one  obtains data that spans  as wide a range  in luminosities as
possible. Furthermore,  even if such  curvature is detected,  there is
little hope  that one can  distinguish between effective  feedback and
ineffective cooling, both of which leave similar features in the slope
and  scatter  of  the  TFR.   Although our  models  are  probably  not
completely realistic, this suggests  that the TFR does not necessarily
provide the most useful constraints on disc galaxy formation.

Another approach that has been taken  in the past, is to use published
luminosity  functions and  luminosity-velocity relations  to construct
halo velocity  functions (i.e., Shimasaku 1993; Newman  \& Davis 2000;
Gonzales \etal  2000; Bullock \etal  2001a; Kochanek 2001).   The main
goal of these studies is similar to the work presented here, namely to
circumvent the problems with poorly understood astrophysical processes
when linking  the observed  properties of galaxies  to those  of their
dark matter haloes.  Our results imply  that great care is to be taken
in  linking  an observable  velocity  such  as  $V_{\rm max}$  to  the
circular  velocity of  a dark  matter  halo (see  also discussions  in
Gonzales  \etal 2000  and Kochanek  2001).  Based  on our  results, we
suggest  that the  construction of  a halo  {\it mass}  function using
$M_{\rm vir} \propto R_d V^2_{\rm max}$ may proof more reliable.
 
In the absence of feedback $f_{\rm gal}$ depends only on halo mass and
is independent  of the halo angular momentum.   However, when feedback
is  included $f_{\rm  gal}$  correlates strongly  with  the halo  spin
parameter $\lambda$ (at least  for the less massive systems).  Systems
with  less   angular  momentum  produce  discs   with  higher  surface
brightness, which  have higher star formation  rates, and consequently
also  higher mass  ejection  rates.  Therefore,  if  we could  somehow
determine the halo  spin parameter, we could use  its correlation with
$f_{\rm gal}$  (or absence thereof), to constrain  the efficiencies of
feedback  processes. In  a  recent  study van  den  Bosch, Burkert  \&
Swaters (2001) used  high quality rotation curve data  for a sample of
dwarf  galaxies to  construct $f_{\rm  gal}(\lambda)$.  Interestingly,
they  found a narrow  correlation with  $f_{\rm gal}$  decreasing with
decreasing  $\lambda$, as  predicted by  our models.   Although  it is
tempting  to interpret  this as  evidence that  feedback does  play an
important role  in disc galaxy formation, there  are several problems.
First  of all,  the  slope of  the  `observed' $f_{\rm  gal}(\lambda)$
relation is  steeper than predicted  by the model. Secondly,  the spin
parameter of  the disc  material (which is  the one obtained  from the
observations) is not  necessarily the same as that  of the dark matter
halo. Finally,  there are  various uncertainties related  to obtaining
estimates of both $\lambda$ and  $f_{\rm gal}$ from the rotation curve
analysis  that may  cause systematic  errors. Nevertheless,  the small
amount  of  scatter  in  $f_{\rm  gal}(\lambda)$  in  both  the  model
predictions and  the data suggests that  a similar analysis  as in van
den  Bosch, Burkert \&  Swaters for  a larger  sample of  galaxies may
proof extremely useful for constraining the feedback efficiencies.

The structural properties of  present day disc galaxies (scale length,
disc-to-bulge  ratio, gas mass  fraction, rotation  velocities) depend
mainly  on  mass  and  angular  momentum. They  contain  virtually  no
information  about the  mass accretion  history (MAH)  of  the galaxy,
which  mainly determines the  current and  past star  formation rates.
Therefore, one might  hope to be able to use  the observed cosmic star
formation history  of disc galaxies to  infer the overall  MAH of dark
matter haloes, which can be used to constrain cosmological parameters.
However, as we  have shown, small changes in  the feedback and cooling
efficiencies  impact strongly  on  the star  formation histories.   In
fact, many different combinations  of cosmology, feedback, and cooling
efficiency can produce similar star formation histories, which implies
that the interpretation of the cosmic star formation history is highly
degenerate.

Finally we investigated how  cooling and feedback influence the colors
and metallicities of the model galaxies. In models with feedback, more
luminous systems  are more efficient in retaining  metals, producing a
fairly narrow  metallicity-luminosity relation. Surprisingly,  also in
models  without   feedback  the   more  luminous  galaxies   are  more
enriched. This  owes to the  star formation threshold density,  due to
which more massive systems can  convert a larger fraction of their gas
into  stars, causing  more enrichment.   Contrary to  observations, we
find  color-magnitude   relations  in  which   brighter  galaxies  are
bluer. This is  a consequence of the hierarchical  nature of structure
formation, in  which more massive  structures form later.   Since star
formation  rates are  mainly  driven  by the  rate  at which  galaxies
accrete mass,  more luminous galaxies  end up with  relatively younger
stellar  populations.  Even  though  the more  luminous galaxies  have
substantially higher metallicities (which  tends to make them redder),
the  differences  in stellar  population  ages  dominates, and  causes
brighter galaxies  to be  bluer.  It remains  to be seen  whether more
sophisticated  modeling  of   the  chemical  enrichment  and  feedback
processes, combined with a treatment of dust extinction can solve this
intricate problem.


\section*{Acknowledgements}

I  am  grateful to  Anthony  Brown,  Guinevere  Kauffmann, Houjun  Mo,
Chenggang Shu, and Simon White for suggestions, advice and stimulating
discussions, and to the anonymous referee for insightful comments that
helped to improve the paper.


\label{lastpage}

\end{document}

%% file: macros_vdbosch.tex
\newcommand{\etal}{{et al.~}}

\newcommand{\lta}{\la}
\newcommand{\gta}{\ga}

\newcommand{\kmsmpc}{\>{\rm km}\,{\rm s}^{-1}\,{\rm Mpc}^{-1}}
\newcommand{\kms}{\>{\rm km}\,{\rm s}^{-1}}

\newcommand{\kpc}{\>{\rm kpc}}
\newcommand{\Msun}{\>{\rm M_{\odot}}}
\newcommand{\Lsun}{\>{\rm L_{\odot}}}

\newcommand{\apj}{ApJ}
\newcommand{\apjs}{ApJS}
\newcommand{\aj}{AJ}
\newcommand{\mnras}{MNRAS}
\newcommand{\aap}{A\&A}

\newdimen\hssize
\newdimen\hdsize